\documentclass[nobibnotes,twocolumn,nofootinbib,superscriptaddress]{revtex4-1}
\usepackage{graphics,color,array,dcolumn}
\usepackage{calc}
\usepackage[T1]{fontenc}
\usepackage[latin1]{inputenc}
\usepackage{amsmath}
\usepackage{amssymb}
\usepackage{xspace}
\usepackage{adjustbox}
\usepackage[colorlinks=true, linkcolor=black, citecolor=black, urlcolor=black, filecolor=black]{hyperref}
\allowdisplaybreaks[1]

\def\gn{G_{\rm N}}

\newcommand{\be}{\begin{equation}}
\newcommand{\ee}{\end{equation}}
\newcommand{\ea}{\end{eqnarray}}
\newcommand{\ba}{\begin{eqnarray}}

\makeatother

\begin{document}
\title{Cosmological bounds on the field content of asymptotically safe gravity-matter models}
\author{Alfio Bonanno}
\address{INAF, Osservatorio Astrofisico di Catania, via S. Sofia 78, I-95123 Catania, Italy \\INFN,  Sezione di Catania,  via S. Sofia 64, I-95123, Catania, Italy}
\author{Alessia Platania}
\address{Institut f\"ur Theoretische Physik, Universit\"at Heidelberg, Philosophenweg 16, 69120 Heidelberg, Germany}
\author{Frank Saueressig}
\address{Institute for Mathematics, Astrophysics and Particle Physics (IMAPP),\\ Radboud University Nijmegen, Heyendaalseweg 135, 6525 AJ Nijmegen, The Netherlands}

\begin{abstract}
We use the non-Gaussian fixed points (NGFPs) appearing in the renormalization group flow of gravity and gravity-matter systems to construct models of NGFP-driven inflation via a renormalization group  improvement scheme. The cosmological predictions of these models depend sensitively on the characteristic properties of the NGFPs, including their position and stability coefficients, which in turn are determined by the field content of the underlying matter sector. We demonstrate that the NGFPs appearing in gravity-matter systems where the matter content is close to the one of the standard model of particle physics are the ones compatible with cosmological data. Somewhat counterintuitively, the negative fixed point value of the dimensionless cosmological constant is essential for these findings.
\end{abstract} 
\maketitle

\section{Introduction}

The Planck and WMAP satellite missions have measured the fluctuation spectrum of the cosmic microwave background (CMB) at an hitherto unprecedented precision \cite{pla15:1,pla15:2}. Combined with other cosmological observations like baryon acoustic oscillations obtained from large scale galaxy surveys this data allows to determine cosmic parameters describing the evolution of the early universe rather accurately. In particular the spectral index $n_s = 0.968 \pm 0.006$, the normalization of the scalar power spectrum $\ln 10^{10} \mathcal{A}_* = 3.062 \pm 0.029$, and the bound on the tensor-to-scalar ratio $r < 0.11$ (95 \% CL), reported in \cite{Patrignani:2016xqp}, are determined with a precision where one can actually discriminate between (and also rule out) various inflationary models \cite{Ade:2015lrj}.

A quite surprising outcome of these investigations is that data seems to favor simplicity. Inflationary models based on a single scalar field or modified gravity theories based on an $R^2$-Lagrangian (Starobinsky-inflation) are quite successful in explaining the observation. In particular, the rather simple action
\be\label{Starobinsky:action}
S = \frac{M_{\rm Pl}^2}{2} \int d^4x \sqrt{-g} \left( R + \frac{1}{6m^2}R^2 \right) 
\ee  
predicts a scalar spectral index $n_s$ and a scalar-to-tensor ratio $r$
\be \label{cc:staro}
n_s \simeq 1 - \frac{2}{N} \, , \qquad r \simeq \frac{12}{N^2}
\ee
where $N$ is the number of $e$-folds of expansion between the creation of the fluctuation spectrum and the end of inflation. For typical values of $N$ between 50 and 60, this is in striking agreement with observations \cite{Ade:2015lrj}. Moreover, the normalization of the scalar power spectrum indicates that the scalaron mass $m$ should be of the order $m^2 \sim 10^{-10} M_{\rm Pl}^2$ \cite{pla15:2,inflationaris}.

In this work we study a close relative of Starobinsky inflation, the so-called NGFP-driven inflation. This scenario is motivated by the idea of Weinberg \cite{1976W,we79} that the short distance behavior of gravity may be controlled by a non-Gaussian fixed point (NGFP) of the gravitational renormalization group flow. In this framework coupling constants like Newton's constant $G$ or the cosmological constant $\Lambda$ become energy-scale dependent. At energies above the Planck scale the ``running'' of these couplings is controlled by the NGFP which ensures that their dimensionless counterparts $g_k = G_k k^2$, $\lambda_k = \Lambda_k k^{-2}$, $\ldots$ approach constant values as the energy scale $k$ goes to infinity. In this way the quantum theory is free from unphysical UV-divergences or, synonymously, ``asymptotically safe''. Starting from the seminal work \cite{Reuter:1996cp} this scenario has been widely explored for pure gravity \cite{Reuter:1996cp,dou,so99,la02,resa02b,resa02,daniel04,bore05,co09,bema10,
mare11,ast13,beca,bogu,begu14,begu,demme,dimo15,Souma:1999at,Codello:2007bd,Machado:2007ea,Falls:2013bv,ast15,Falls:2016msz,Dietz:2012ic,Ohta:2015efa,Ohta:2015fcu,Codello:2006in,Benedetti:2009rx,Groh:2011vn,Gies:2016con,Rechenberger:2012pm,Christiansen:2012rx,Manrique:2011jc, AleFrank1, AleFrank2,Platania:2017djo,2017franktopos} and, more recently, also in the context of gravity-matter systems \cite{Percacci:2002ie,Percacci:2003jz,Zanusso:2009bs,Vacca:2010mj,Harst:2011zx,Eichhorn:2011pc,Dona:2012am,Dona:2013qba,Dona:2014pla,peva14,Meibohm:2015twa,Oda:2015sma,Dona:2015tnf,Labus:2015ska,Meibohm:2016mkp,Eichhorn:2016esv,Eichhorn:2017ylw,Alkofer:2018fxj}. This led to the key insight that gravity as well as many phenomenologically interesting gravity-matter systems possess a NGFP suitable for realizing asymptotic safety.

While this scenario is quite attractive from a quantum field theory point of view, since it allows for a consistent quantization of the gravitational force along the same lines of the other fundamental interactions, it also comes with the technical challenge that the action describing the gravitational interactions in the high-energy phase is quite complicated and not well understood. It is clear, however, that it contains higher curvature terms similar to the ones encountered in \eqref{Starobinsky:action}. Moreover, one expects that the set of asymptotically safe theories can be parameterized by a  finite number of free parameters which need to be fixed based on experimental input.

Our present work then builds on two recent observations: First, it has been shown in \cite{AlfioAle1} that, at the level of pure gravity, the class of Asymptotically Safe actions can be tested using cosmological data\footnote{For other works connecting Asymptotic Safety to cosmology see the original works \cite{irfp,br02,dou,guberina03,2004reuterw1,2004reuterw2,
		guberina05,resa05,boes,br07,weinberg10,cai11,conpe,alfio12,
		AlfioAle1,cai13,2012PhRvD..85d3501C,2016JCAP...02..048S,2017JCAP...07..015T} and \cite{2017alfr} for a review.}. In particular it has been found that Starobinsky-type actions appear quite naturally, indicating that one may obtain interesting cosmologies within this set of models. Second, it has been understood that the addition of matter fields may give rise to NGFPs whose properties differ from the one found in the case of pure gravity. In particular their position projected to the $g$-$\lambda$--plane and the value of the stability coefficients, governing the scale-dependence of the couplings in the high-energy regime, depend on the precise matter content of the system. In \cite{AleFrank2} it was demonstrated that one can distinguish three broad classes of gravity-matter fixed points according to their stability properties: Class I, represented by pure gravity, has a spiraling attractor, Class II, containing the standard model of particle physics, has real critical exponents and a negative product $g_\ast\lambda_\ast$ while Class III, comprising GUT-type models, comes with real critical exponents and a positive product $g_\ast\lambda_\ast$, see Table \ref{Tab.3}.\footnote{The link between the number of matter fields and the corresponding fixed point classification actually depends on the regularization scheme. Depending on the computational details like, e.g.\ the parameterization of the fluctuation fields, not all classes may be realized within a given setting. The conclusions of this work can then be drawn at the level of the fixed point properties, independently of the precise set of matter fields realizing the corresponding class.}
\begin{table*}[t!]
\renewcommand{\arraystretch}{1.4}
\begin{center}
\begin{adjustbox}{width=0.98\textwidth}
\begin{tabular}{|p{3.4cm}||c|c|c||c|c|c|c|c|c||c|c|c|c|c|c|} \hline \hline
Model & \multicolumn{3}{c||}{Field content} & \multicolumn{6}{c||}{Foliated} & \multicolumn{6}{c|}{Covariant}\\ \hline
& $N_S$ & $N_D$ & $N_V$ & $g_\ast$ & $\lambda_\ast$ & $\theta_1$ & $\theta_2$ & $e_{1}^{2}$ & $e_{2}^{1}$ &  $g_\ast$ & $\lambda_\ast$ & $\theta_1$ & $\theta_2$ & $e_{1}^{2}$ & $e_{2}^{1}$  \\\hline\hline
Pure gravity & 0 & 0 & 0 & $0.78$ & $+\,0.32$ & \multicolumn{2}{c|}{$0.503 \pm 5.387 \, i$} & $0$ & $0$ & 
$2.53$ & $0.18$ & 4 & 2.775 & 0 & 0.22 \\ \hline
Standard Model (SM) & 4 & $\tfrac{45}{2}$ & 12 & $0.75$ & $-\,0.93$ & 3.871 & 2.057 & -0.06 & -1.26 &   
$0.54$ & -0.63 & 4 & 2.127 & 0 & -1.09
\\ \hline
SM, dark matter (dm) & 5 & $\tfrac{45}{2}$ & 12 & $0.76$ & $-\,0.94$ & 3.869 & 2.058 & -0.07 & -1.24 & 
$0.55$ & -0.63 & 4 & 2.129 & 0 & -1.07
\\ \hline
SM, $3\,\nu$ & 4 & 24 & 12 & $0.72$ & $-\,0.99$ & 3.884 & 2.057 & -0.06 & -1.38 & 
$0.52$ & -0.66 & 4 & 2.121 & 0 & -1.21
\\ \hline
SM, $3\,\nu$, dm, axion & 6 & 24 & 12 & $0.75$ & $-\,1.00$ & 3.882 & 2.059 & -0.05 & -1.34 & 
$0.54$ & - 0.66 & 4 & 2.126 & 0 & -1.16
\\ \hline
MSSM (SM Higgs) & 49 & $\tfrac{61}{2}$ & 12 & $2.26$ & $-\,2.30$ & 3.911 & 2.154 & -0.16 & -1.01 & 
$1.27$ & -1.15 & 4 & 2.327 & 0 & -0.81
 \\ \hline
MSSM (Higgs doublet) & 53 & $\tfrac{63}{2}$ & 12 & $2.66$ & $-2.72$ & 3.924 & 2.162 & -0.16 & -1.01 & 
$1.41$ & -1.28 & 4 & 2.370 & 0 & -0.81
 \\ \hline
{SU(5) GUT} & {124} & {24} & {24} & $0.17$ & $+\,0.41$ & 25.26 & 6.008 & -0.05 & 1.48  & $-$ & $-$ & $-$ & $-$ & $-$ & $-$  \\ \hline
{SO(10) GUT} & {97} & {24} & {45} & $0.15$ & $+\,0.40$ & 19.20 & 6.010 & -0.27 & 1.80 & $-$ & $-$ & $-$ & $-$ & $-$ & $-$ \\ \hline \hline
\end{tabular}
\end{adjustbox}
\end{center}
\caption{\label{Tab.3} Fixed point structure arising from gravity-matter systems on foliated spacetimes \cite{AleFrank2} (foliated columns) and the $f(R)$-computation \cite{Alkofer:2018fxj} (covariant columns). {In the foliated case the gravitational degrees of freedom are carried by the fields appearing in the Arnowitt-Deser-Misner formalism \cite{Arnowitt:1962hi} while the covariant computation considers fluctuations of a graviton $h_{\mu\nu}$ in a fixed background. Despite the different ways of encoding the gravitational degrees of freedom, both settings give rise to remarkably similar values for the critical exponents $\theta_i$. The Standard Model (SM), its minor modifications, and the Minimal Supersymmetric Standard Model (MSSM) induce a gravity-matter NGFP with $\lambda_\ast g_\ast<0$. The difference in the fixed point structure found for GUT-type models can be traced back to the different gauge-fixing procedures employed by the computations (see \cite{Alkofer:2018fxj} for a detailed discussion of this point).}  The eigenvectors ${\bf e}_{1,2}$ are normalized such that $e_{1}^{1} = e_{2}^{2} = 1$.}
\end{table*}

In this work we construct NGFP-driven inflationary models based on the three characteristic classes of non-Gaussian gravity-matter fixed point and compare the resulting cosmological predictions. It thereby turns out that the inflationary models based on the Class II fixed points are highly favored. Counterintuitively, it is the negative value of the dimensionless cosmological constant at the NGFP which makes these models viable.\footnote{Note that a negative cosmological constant in the NGFP-regime is not in conflict with the positive value of $\Lambda$ observed at cosmological scales: the NGFP-regime and the classical regimes are connected through a crossover in which the sign of $\Lambda_k$ changes at energies around the Planck scale.} In this way the interplay between cosmological measurements and NGFP-driven inflation puts (rather mild) bounds on the matter fields present in Nature.

The rest of the work is organized as follows. In Section \ref{sect.2} we briefly review the properties of the NGFPs encountered in gravity-matter systems following \cite{AleFrank2,Alkofer:2018fxj}. The inflationary models arising from these fixed points are constructed in Section \ref{AFA} and the resulting cosmological signatures are derived in Section \ref{sect.4}. The consequences of our findings are discussed in Section \ref{conclu}.

\section{Asymptotic Safety in a nutshell}
\label{sect.2}
A general feature of a quantum field theory is that its coupling constants $u^\alpha$ depend on the energy scale $k$. In the vicinity of a fixed point, located at $u_*^\alpha$, the full RG flow is captured by the linearized beta functions
\be\label{flow:lin}
k \partial_k u^\alpha = \sum_{\gamma} B^\alpha{}_\gamma (u^\gamma_k - u_\ast^\gamma) 
\ee
where the components
of the stability matrix $B$ are given by $B^\alpha{}_\gamma \equiv \partial_{u^\gamma} \beta^\alpha |_{u = u_\ast}$. Denoting the 
 right-eigenvectors of $B$ by $e^\alpha_i = ({\bf e}_i)^\alpha$ the stability coefficients $\theta_i$ of the system are defined as minus the eigenvalues of $B$, i.e.\ $ \sum_{\gamma} B^\alpha{}_\gamma e^\gamma_i = - \theta_i e^\alpha_i$. The solution of the linearized equation \eqref{flow:lin} is then given by
 \be\label{eq:linsol}
 u^\alpha(k) = u_\ast^\alpha + \sum_i c_i \, e_i^\alpha \left( \frac{k}{k_0} \right)^{-\theta_i} \, ,
 \ee
where $c_i$ are free integration constants and $k_0$ is an arbitrary renormalization scale. The asymptotic safety condition, stating that the solution is attracted to the fixed point as $k\rightarrow \infty$, then entails that the coefficients $c_i$ associated to the stability coefficients with Re$\,\theta_i < 0$ must be fixed to zero. Based on the construction of the complete phase diagram \cite{resa02,Christiansen:2014raa}, one concludes that the linearized solution \eqref{eq:linsol} provides a good approximation at energies $k^2 \gtrsim M_{\rm Pl}^2$.

For the gravity-matter flows projected to the Einstein-Hilbert action, where $u^\alpha_k = \{\lambda_k, g_k\}$, it is actually useful to distinguish between the two cases where $\theta_1$ and $\theta_2$ are real (Classes II and III) or form a complex conjugate pair {(Class I)}. In the first case \eqref{eq:linsol} entails
\begin{subequations}  \label{scalingrc}
\begin{align} \label{eq5a}
\lambda_k = \lambda_\ast &+ c_1 \,e_{1}^{1} \, \big(\tfrac{k}{k_0}\big)^{-\theta_1}+c_2\, e_{2}^{1} \, \big(\tfrac{k}{k_0}\big)^{-\theta_2} \;\;, \\
g_k =g_\ast &+c_1 \, e_{1}^{2} \, \big(\tfrac{k}{k_0}\big)^{-\theta_1}+c_2 \, e_{2}^{2} \, \big(\tfrac{k}{k_0}\big)^{-\theta_2} \;\; . \label{ggsc}
\end{align}
\end{subequations}
The only free parameters are the dimensionless variables $c_1$ and $c_2$ which allow to select a particular RG trajectory and can be determined by comparing physical observables, e.g.\ physical couplings in the effective Lagrangian, with observations. 

For complex stability coefficients, $\theta_{1,2} = \theta^\prime \pm i \theta^{\prime\prime}$, the eigenvectors ${\bf e}_{1,2}$ are complex conjugates of each other. Redefining ${\bf e}_1 = {\rm Re} \, {\bf e}_1$ and ${\bf e}_2 = {\rm Im} \, {\bf e}_2$ and similarly for $c_{1,2}$ the general solution can be written as
\begin{subequations}  \label{scalingrc2}
\begin{align}
\lambda_k = & \lambda_\ast  + \, \Big( \left( c_1 \cos(\theta^{\prime\prime} t) + c_2 \sin(\theta^{\prime\prime} t)   \right) e_1^1 \\ \nonumber
& + \left( c_1 \sin(\theta^{\prime\prime} t) - c_2 \cos(\theta^{\prime\prime} t)  \right) e_2^1 \Big) 
\left( \frac{k}{k_0} \right)^{-\theta^\prime} \;\;, \\ \label{ggsc2}
g_k = & g_\ast  + \, \Big( \left( c_1 \cos(\theta^{\prime\prime} t) + c_2 \sin(\theta^{\prime\prime} t)   \right) e_1^2 \\ \nonumber
& + \left( c_1 \sin(\theta^{\prime\prime} t) - c_2 \cos(\theta^{\prime\prime} t)  \right) e_2^2 \Big) 
\left( \frac{k}{k_0} \right)^{-\theta^\prime} \;\;. 
\end{align}
\end{subequations}
Here $t \equiv \ln(k/k_0)$. 

For later reference, we note the inversion formula for $g_k$. Employing the geometric series expansion, eq.\ \eqref{ggsc} gives
\be\label{ginvreal}
\tfrac{1}{g_k} = \tfrac{1}{g_*} \sum_{n=0}^\infty (-1)^n \left[ \tfrac{c_1}{g_*} e_1^2 \big(\tfrac{k}{k_0}\big)^{-\theta_1}+ \tfrac{c_2}{g_*} \, e_{2}^{2} \, \big(\tfrac{k}{k_0}\big)^{-\theta_2} \right]^n \, . 
\ee
Ordering the stability coefficients according to $\theta_1 \ge \theta_2 > 0$, and assuming that $k/k_0 \gg 1$, the leading terms in this expansion are
\be
\frac{1}{g_k} \simeq \frac{1}{g_*} \left[ 1 -  \tfrac{c_2}{g_*} \, e_{2}^{2} \, \big(\tfrac{k}{k_0}\big)^{-\theta_2} + \mbox{subleading} \right] \, . 
\ee
Analogously, the case of complex critical exponents \eqref{ggsc2} leads to
\be
\frac{1}{g_k} = \frac{1}{g_*} \sum_{n=0}^\infty (-1)^n \left[ \frac{1}{g_*} \left( \ldots \right) \left( \frac{k}{k_0} \right)^{-\theta^\prime} \right]^n \, , 
\ee
where the dots represent the expression in the round bracket. 

We close the section by introducing ``reference models'' for each of the three classes represented in Table \ref{Tab.3}.
\begin{subequations}
\begin{align} \label{model1}
&\mbox{Class \, I}: && \quad\;\;
\begin{array}{ll}
\theta^\prime = 0.5 \, , & \qquad \;\; {\bf e}_1 = \{1,0\}   \\[1.1ex]
\theta^{\prime\prime} = 5 \, , & \qquad \;\; {\bf e}_2 = \{0,1\}
\end{array} 
\\[1.1ex]
\label{model2}
&\mbox{Class \, IIa}: && \quad\;\;
\begin{array}{ll}
\theta_1 = 4 \, , & \qquad \,\;\;\;\; {\bf e}_1 = \{1,0\}   \\[1.1ex]
\theta_2 = 2 \, , & \qquad \,\;\; \;\; {\bf e}_2 = \{-1,1\}
\end{array}
\\[1.1ex]
\label{model2b}
&\mbox{Class \, IIb}: && \quad\;\;
\begin{array}{ll}
	\theta_1 = 4 \, , & \qquad \,\;\;\;\; {\bf e}_1 = \{1,0\}   \\[1.1ex]
	\theta_2 = 3 \, , & \qquad \,\;\;\;\; {\bf e}_2 = \{0.2,1\}
\end{array}
\\[1.1ex]
\label{model3}
&\mbox{Class \, III}: && \quad\;\;
\begin{array}{ll}
\theta_1 = 25 \, , & \qquad \;\;\; {\bf e}_1 = \{1,0\}   \\[1.1ex]
\theta_2 = 6 \, , & \qquad \;\;\; {\bf e}_2 = \{1.5,1\} 
\end{array} 
\end{align}
\end{subequations}
{These reference models mainly serve the purpose of highlighting the qualitatively different cosmological dynamics which results from the RG improved actions constructed below. Apart from the critical exponents listed for pure gravity (Class I), the values for $\theta_i$ correspond to ``typical values'' for critical exponents obtained in the corresponding setting.\footnote{The precise values for the critical exponents associated with the NGFPs appearing in pure gravity and gravity-matter systems are currently not known. The values listed above reflect current ``best estimates'' which may vary slightly depending on the precise form of the approximations made when evaluating the functional renormalization group equation, choice of regularization procedure, or spacetime signature \cite{Manrique:2011jc}.} In the case of pure gravity, a detailed comparison \cite{AleFrank1} of critical exponents obtained from different computations shows that $\theta^\prime \simeq 1.5$ and $\theta^{\prime\prime} \simeq 3$ may be a better approximation to the exact value. Notably, the cosmological analysis carried out below is actually independent of the way the critical exponents are computed and the cosmological constraints remain valid independently of the gravity-matter system which realizes the corresponding critical exponents.}

{At this stage the following remark is in order. The linearized approximation of the RG flow \eqref{flow:lin} holds in the vicinity of a RG fixed point only. The full phase diagrams for selected gravity-matter models have been constructed in \cite{AleFrank2}. These show that even in the case where $\lambda_\ast \le 0$, there are RG trajectories which undergo a crossover to $\lambda_k > 0$ at observable scales. Thus $\lambda_\ast \le 0$ is not in tension with the positive value of the cosmological constant observed at cosmic scales.}

\section{Effective Lagrangians for quantum gravity-matter systems}
\label{AFA}

In the present discussion we restrict the gravitational sector to the Einstein-Hilbert truncation. The corresponding scale-dependent Lagrangian reads
\be\label{runningEH}
\mathcal{L}_k=\frac{k^2}{16 \pi g_k}(R-2\lambda_k k^2)  \, .
\ee
Throughout the study, we assume that inflation is driven by quantum gravitational effects so that the dynamics of the matter sector can be neglected.

A RG-improved Lagrangian can be obtained by substituting the running couplings \eqref{scalingrc} into eq.~\eqref{runningEH} \cite{fayos11,br00,br07,Reuter:2004nx,boes,2004reuterw1,2004reuterw2,alfio12}. The resulting analytical expression strongly depends on the critical exponents $\theta_i$. In particular, following the discussion in \cite{AleFrank2}, these stability coefficients are complex conjugate in the case of pure gravity case, positive and real for gravity-matter systems such as the Standard Model (SM) and its modifications (see Table~\ref{Tab.3}).

In order to close the system, the infrared cutoff $k$ must be related to a diffeomorphism-invariant quantity such that the resulting semi-classical model is described by an effective diffeomorphism-invariant Lagrangian.\footnote{In the case of scalar-field theory a similar cutoff-identification allows to retrieve the Coleman-Weinberg potential in a rather straightforward manner \cite{cw,1973migdal,1973gross,1978pagels,1978Matinyan,1983adler}.}
The simplest scale-setting satisfying these requirements is \cite{Dietz:2012ic,alfio12,saltas12,co15}
\be\label{scalesetting}
 k^2= \xi R  \, . 
\ee 
{Here $\xi$ is an a priori undetermined, positive constant whose value may be fixed by comparing to a one-loop computation \cite{Koch:2014cqa} or imposing other physical requirements.} 
The cutoff identification \eqref{scalesetting} generates an effective $f(R)$-type Lagrangian in a natural way. Notably, evaluating eq.~\eqref{runningEH} at the fixed point, where $g_k = g_\ast$ and $\lambda_k = \lambda_\ast$, the scale-setting \eqref{scalesetting} gives rise to the $R^2$-type action identified in \cite{beca,2013morrisdietz,2015demsau} as the RG fixed point of $f_k(R)$-gravity.

Starting from the scale-dependent Lagrangian \eqref{runningEH}, substituting the scale-dependent couplings from eqs.\ \eqref{eq5a} and \eqref{ginvreal} (truncating the geometric expansions at $n=1$) and applying the scale-setting procedure \eqref{scalesetting} results in the following RG-improved Lagrangian
\begin{align}
\mathcal{L}_\mathrm{eff} &= a_0 \, R^2 + b_1\,R^{\frac{4-\theta_1-\theta_2}{2}} + b_2\,R^{\frac{4-\theta_1}{2}} \notag \\
& + b_3\,R^{\frac{4-\theta_2}{2}}+b_4\,R^{2-\theta_1} + {b_5}\,R^{2-\theta_2}\;\;. \label{genRGlag}
\end{align}
The coefficient of the $R^2$-term is given by
\be\label{a0coeff}
a_0=\frac{\xi\,(1-2 \xi \lambda_\ast)}{16 \pi g_\ast} \; ,
\ee
while the $b_i$'s read
\begin{subequations}
\begin{align}
&b_1=\frac{c_1 c_2 (e_1^1 \,e_2^2{+e_1^2 \,e_2^1})\,\xi^{\frac{4-\theta_1-\theta_2}{2}}\,k_0^{\theta_1+\theta_2}}{8\pi g_\ast^2} \;\;, \\
&b_2=\frac{c_1\, ( 2e_1^2\xi\, \lambda_\ast-2 e_1^1 \xi \, g_\ast -e_1^2)\,\xi ^{\frac{2-\theta_1}{2}}\, k_0^{\theta_1}}{16 \pi  g_\ast^2} \;\;, \\
&b_3=\frac{c_2\, ( 2e_2^2\xi\, \lambda_\ast-2 e_2^1 \xi \, g_\ast -e_2^2)\,\xi^{\frac{2-\theta_2}{2}}\,k_0^{\theta_2}}{16\pi g_\ast^2} \;\;, \\
&b_4= \frac{c_1^2\, (e_1^1 \, e_1^2)\,\xi^{{2-\theta_1}}\,k_0^{2\theta_1}}{8\pi g_\ast^2}\;\;, \\
&b_5=\frac{c_2^2\, (e_2^1 \, e_2^2)\,\xi^{{2-\theta_2}}\,k_0^{2\theta_2}}{8\pi g_\ast^2} \;\;.
\end{align}
\end{subequations}
Notably, the scale-setting procedure always produces an $R^2$-term whose coefficient is determined by $(\lambda_\ast,g_\ast,\xi)$. The occurrence of this term is universal in the sense that it does not depend on the field content of the model and solely relies on the presence of a NGFP. The information about the stability coefficients $\theta_i$ and the chosen RG trajectory ($c_i$-coefficients) is carried by the $b_i$-terms. As it will turn out, it is these terms that encode the subleading corrections to the scale-free fluctuation spectrum. Also note that negative stability coefficients $\theta_i < 0$, which would naively dominate \eqref{genRGlag} in the large curvature regime, do not enter $\mathcal{L}_\mathrm{eff}$. From eq.\ \eqref{eq:linsol} one sees that such terms come with positive powers of $k$, so that the corresponding coefficients $c_i$ must be set to zero to ensure that the flow approaches the fixed point as $k \rightarrow \infty$.

Following the same steps for a NGFP with complex critical exponents results in a class of improved Lagrangians studied in \cite{alfio12}
\begin{align}\nonumber
\mathcal{L}_\mathrm{eff} &= \, a_0 \, R^2 + \tilde{b}_1 \, R^{\frac{4-\theta^\prime}{2}} \cos\left(\tfrac{1}{2} \,  \theta^{\prime\prime}\ln\left(\tfrac{\xi R}{k_0^2}\right)\right) \\  \label{Leffcc}
& \,+\, \tilde{b}_2 \, R^{\frac{4-\theta^\prime}{2}} \sin\left(\tfrac{1}{2} \,  \theta^{\prime\prime} \ln\left(\tfrac{\xi R}{k_0^2}\right)\right) \\
& \, +\,\tilde{b}_3 \, R^{{2-\theta^\prime}}+ \tilde{b}_4 \, R^{{2-\theta^\prime}} \cos\left(\theta^{\prime\prime}\ln\left(\tfrac{\xi R}{k_0^2}\right)\right) \nonumber\\
& \, +\, \tilde{b}_5 \, R^{{2-\theta^\prime}} \sin\left(\theta^{\prime\prime}\ln\left(\tfrac{\xi R}{k_0^2}\right)\right) \nonumber
\end{align}
where
\begin{subequations}
\begin{align}
\tilde{b}_1 = & \, \tfrac{\left( (2 \xi \lambda_*-1) (c_1 e^2_1 - c_2 e^2_2) - 2 \xi g_* (c_1 e^1_1 - c_2 e^1_2)  \right)\,\xi^{\frac{2-\theta^\prime}{2}}\,k_0^{\theta^\prime}}{16 \pi g_*^2} , \\
\tilde{b}_2 = & \, \tfrac{\left( (2 \xi \lambda_*-1) (c_2 e^2_1 + c_1 e^2_2) - 2 \xi g_* (c_2 e^1_1 + c_1 e^1_2)  \right)\,\xi^{\frac{2-\theta^\prime}{2}}\,k_0^{\theta^\prime}}{16 \pi g_*^2} , \\
\tilde{b}_3 = & \, \tfrac{\left(c_1^2+c_2^2\right)\left(e_1^1 e_1^2+e_2^1 e_2^2\right)\,\xi ^{2-\theta^{\prime}} k_0^{2\theta^{\prime}}}{16\pi g_\ast^2} ,\\ 
\tilde{b}_4 = & \, \tfrac{\left((c_1^2-c_2^2) (e_1^1 e_1^2-e_2^1 e_2^2)-2c_1c_2 (e_1^1 e_2^2 +e_2^1 e_1^2)\right)\,\xi ^{2-\theta^{\prime}} k_0^{2\theta^{\prime}}}{16\pi g_\ast^2},\\ 
\tilde{b}_5 = & \, \tfrac{\left((c_1^2-c_2^2)  (e_1^1 e_2^2 +e_2^1 e_1^2)+2c_1c_2(e_1^1 e_1^2-e_2^1 e_2^2)\right)\,\xi ^{2-\theta^{\prime}} k_0^{2\theta^{\prime}}}{16\pi g_\ast^2}.
\end{align}
\end{subequations}

At this stage, it is instructive to give the explicit form of $\mathcal{L}_\mathrm{eff}$ arising from the prototypical fixed point of Class IIa associated to a typical standard model like matter content. Evaluating the general Lagrangian \eqref{genRGlag} for the fixed point characteristics \eqref{model2} leads to a \emph{Starobinsky-type action} of the form\footnote{Here we drop terms containing inverse powers of $R$ since they do not affect the inflationary dynamics arising from actions of the type \eqref{saction}.}
\be\label{saction}
\mathcal{L}_\mathrm{eff} = \frac{1}{16 \pi \gn} \left( R + \frac{1}{6m^2} R^2 - 2 \Lambda_\mathrm{eff} \ \right) \, . 
\ee
The effective couplings $G_N, m^2$ and $\Lambda_\mathrm{eff}$ in this Lagrangian are determined by the fixed point data $g_*, \lambda_*$ and 
the specific RG trajectory underlying the improvement procedure $c_1, c_2, \xi$. Explicitly,
\begin{subequations}\label{eq19}
\begin{align}
\gn&=\frac{g_\ast^2}{c_2\,(2g_\ast\xi+2\lambda_\ast \xi-1)\, k_0^2} \;\;, \label{k1cond} \\
m^2&=\frac{g_\ast}{6\,\xi\, (1-2\lambda_\ast\xi)\,\gn\,} \;\;, \label{k2cond} \\
\Lambda_\mathrm{eff} & =\frac{k_0^4 \left(c_2^2+{c_1} g_\ast\right)\gn}{g_\ast^2}\,\,. \label{k3cond}
\end{align}
\end{subequations}

\section{Cosmic parameters}
\label{sect.4}
We are now in the position to verify if and under which conditions NGFP-driven inflation gives rise to an inflationary scenario in agreement with the Planck data \cite{pla15:1,pla15:2}. Depending on the critical exponents, some of the $b_i$-contributions to the Lagrangian \eqref{genRGlag} could be in the form $R^{-p}$, with $p$ positive: these terms are suppressed when $R$ is large and become important only on cosmic scales \cite{2006PhLB:capoz}, where the linearized approximation of the RG flow, eq.~\eqref{eq:linsol}, is no longer valid. In particular, if all critical exponents satisfy the condition $\theta_i>4$  then all $b_i$-terms are suppressed \cite{2006PhLB:capoz} and the resulting Lagrangian $\mathcal{L}\sim R^2$ gives rise to a scale invariant scalar power spectrum, $n_s=1$. Therefore,
in order to obtain a phenomenologically interesting inflationary scenario compatible with observational data, at least one critical exponent must be $\theta_i<4$. As $\theta_1>\theta_2$ is assumed, the first constraint we find reads
\be
\theta_2 < 4 \;\;. \label{basicresult1}
\ee

\begin{figure}[t!]
\includegraphics[width=0.5\textwidth]{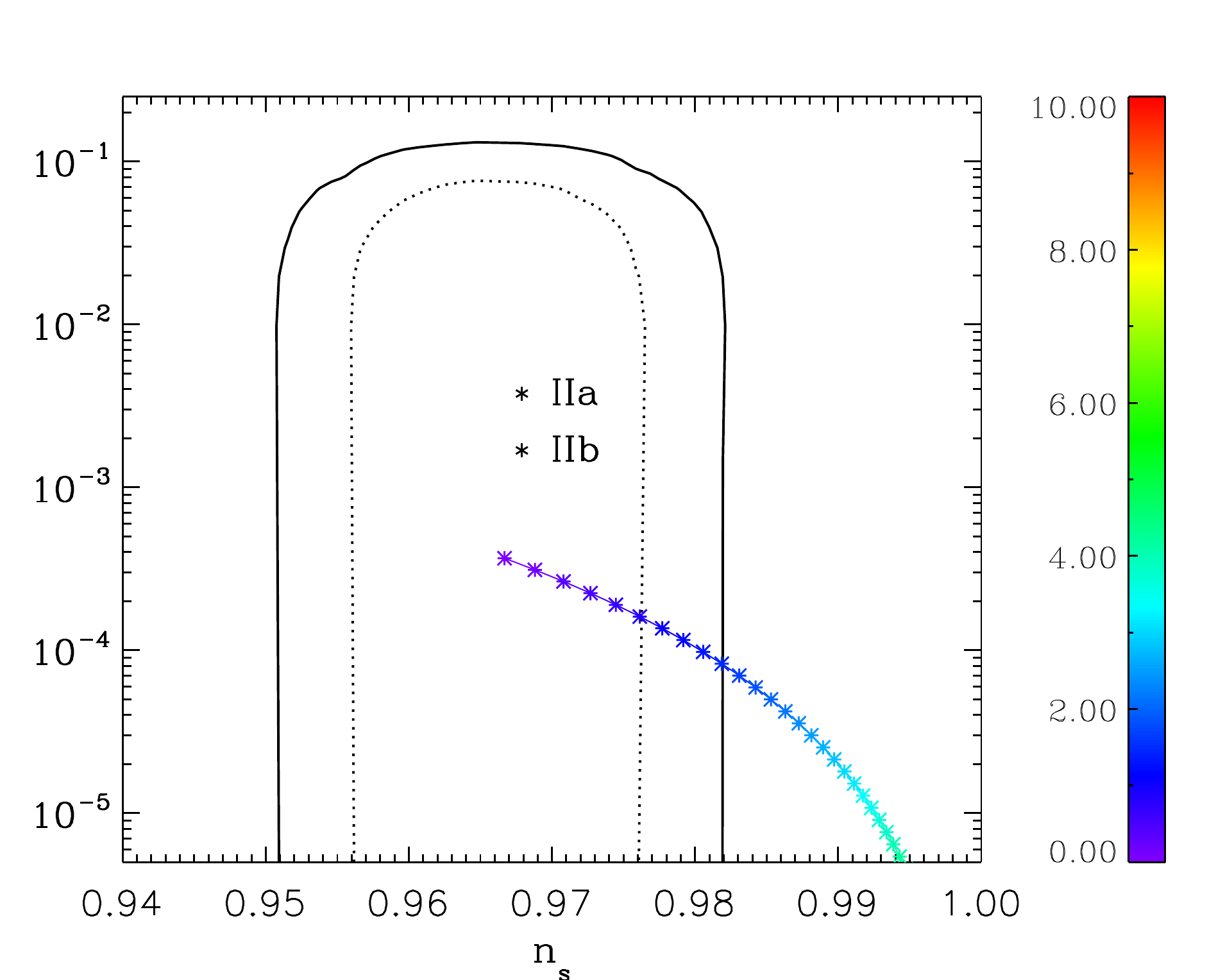}
\caption{Spectral index and tensor-to-scalar ratio induced by the class of theory 
\eqref{GUTaction} as a function of the power index $p$ for $N=60$ e-folds.
Dashed and solid lines represents $1\sigma$ and $2\sigma$ level respectively. The data corresponding to 
Class {{IIa}} and Class {{IIb}} models are also displayed with a $\ast$. 
The values for the cosmic parameters obtained from Class I model are outside the range covered by the figure and therefore not shown. Note that the models IIa and IIb give rise to $r$-values which could be tested by future CMB experiments like CORE \cite{Finelli:2016cyd}, LiteBIRD \cite{Matsumura:2013aja}, PIXIE \cite{Kogut:2011xw}, or CMB S4 \cite{Abazajian:2016yjj}.
 \label{figGUT}}
\end{figure}

\begin{figure}[t!]
\includegraphics[width=0.45\textwidth]{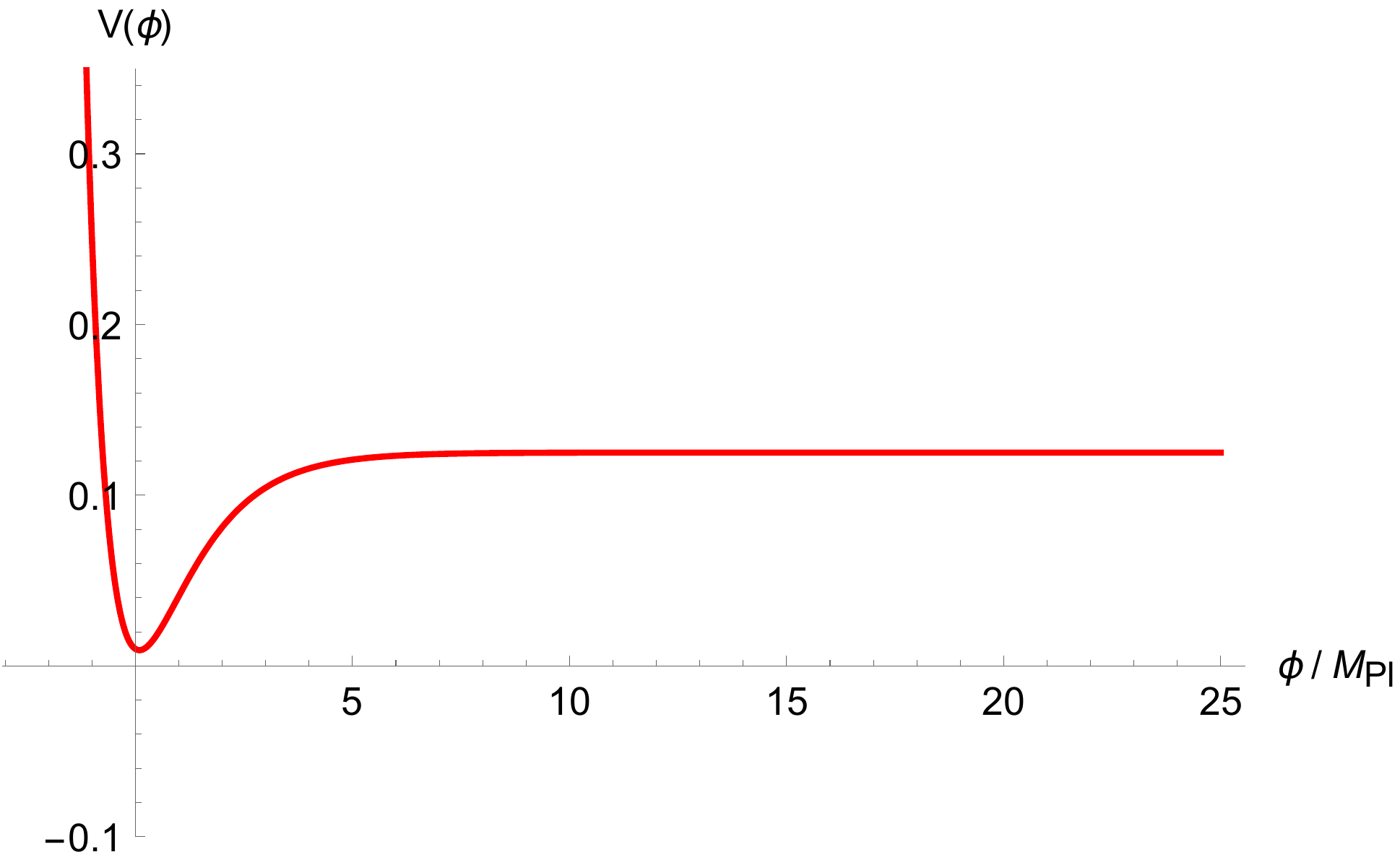}\\[0.5cm]
\includegraphics[width=0.45\textwidth]{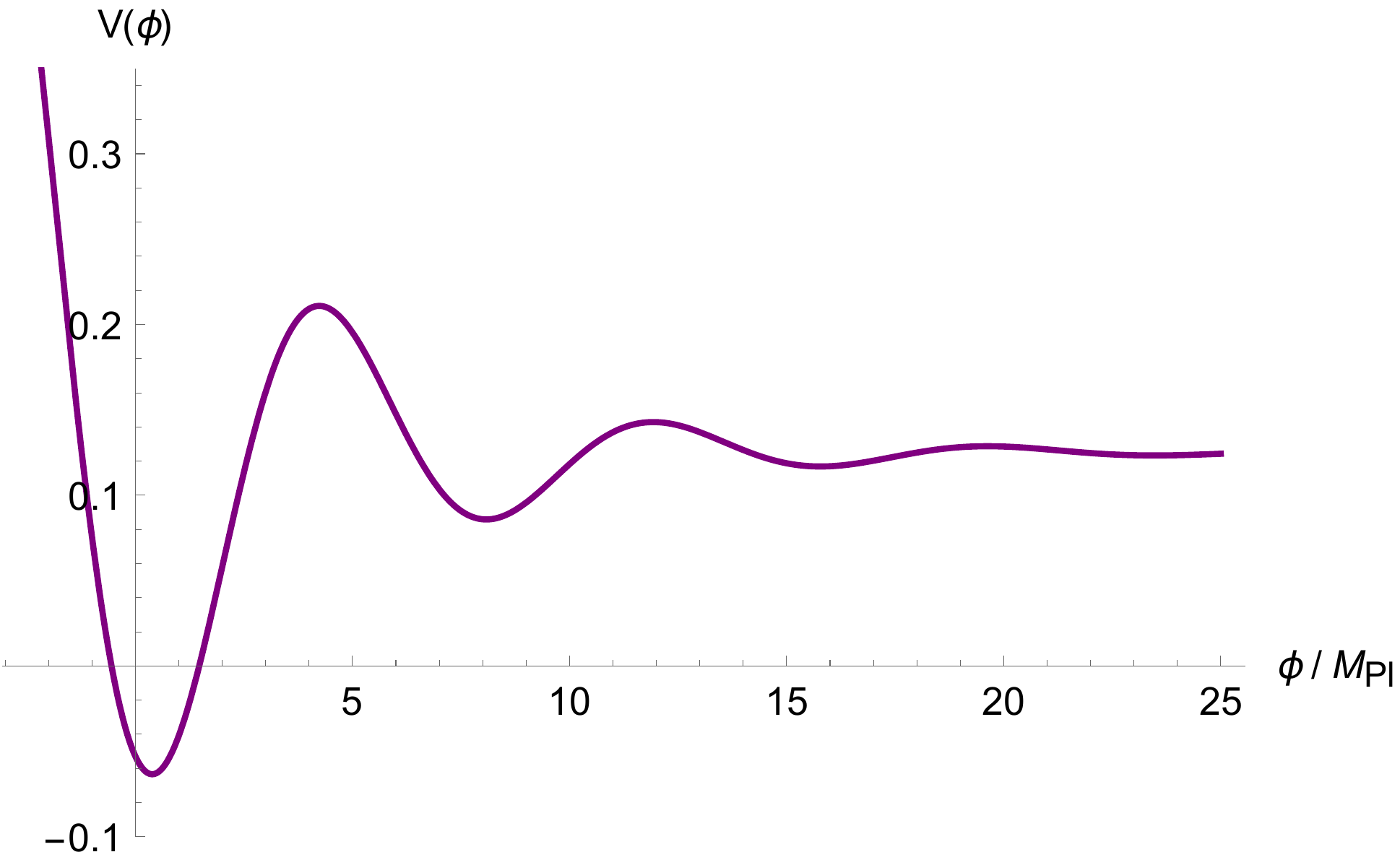}
\caption{Scalar potentials obtained from the action \eqref{saction} (top) and \eqref{Leffcc} evaluated for the fixed point \eqref{model1} and coefficients $\tilde b_1 = \tilde{b}_2 = 1$ and $\tilde b_3 = \tilde{b}_4 = \tilde{b}_5 = 0$ (bottom). In the latter case the potential exhibits an infinite number of metastable de Sitter vacua. An inflationary phase including an exit from the slow roll conditions can be realized by starting close to the maximum at $\phi \simeq 4.23{\,M_\mathrm{Pl}}$ and rolling to the left. \label{fignotGUT}}
\end{figure}

The prototypical NGFPs satisfying this requirement are the ones summarized in Class II. Following Table \ref{Tab.3} these comprise gravity coupled to Standard Model matter and some of its frequently studied beyond the Standard Model extensions. Thus we start by analyzing this class. {In practice this analysis is performed by converting the $f(R)$-type Lagrangian to a scalar-tensor theory by adding suitable Lagrange multipliers (see \cite{defe,cosa15} for a pedagogical account). The resulting scalar potential is illustrated in Fig.\ \ref{fignotGUT}. The results obtained in this way}  are conveniently summarized in Fig.\ \ref{figGUT}.

\medskip
\noindent
{\it NGFPs of Class IIa:} The effective Lagrangian for Class IIa is given in \eqref{saction} with the relation between the NGFP and the effective couplings provided by \eqref{eq19}. Neglecting $\Lambda_\mathrm{eff}$ for a moment, the model reduces to the $R+R^2$-Lagrangian \eqref{Starobinsky:action}. The value of the mass $m$ is fixed by the CMB normalization of the power spectrum, $\mathcal{A}_s\sim10^{-10}$, and for a Starobinsky-like model this implies ${m}\simeq10^{13}\,\text{GeV}$ (see \cite{inflationaris} for details). 
Investigating \eqref{k2cond}, treating $\xi$ as a free parameter, one finds that this condition can only be met if
\be\label{fpcond1}
\lambda_\ast \le 0 \, . 
\ee
Taking $g_*, \lambda_*$ of order unity, the observational constraint 
$\gn m^2 \sim10^{-11}$ fixes $\xi\sim10^5$. The values of $c_1$ and $c_2$ can then be obtained by requiring that the RG trajectory underlying the construction reproduces the values for Newton's constant and the cosmological constant observed at large distances, see \cite{Reuter:2004nx} for a detailed discussion. All parameters appearing in our model are thus completely fixed by observations. Inspecting Table \ref{Tab.3}, it is {quite remarkable that there are indeed gravity-matter fixed points fulfilling the whole set of conditions $\theta_2<4$ and $\lambda_\ast\leq 0$. Even more remarkable, the NGFP found for the matter content of the standard model of particle physics is one representative of this class.} 

We now reinstall the cosmological constant and compute the spectral index $n_s$ and tensor-to-scalar ratio $r$ associated with the class of theories \eqref{saction}. These parameters can be easily computed within the slow-roll approximation \cite{inflationaris}. A straightforward calculation gives
\begin{subequations}
\begin{align}
& n_s =1-\frac{2}{N}+\frac{24\, \Lambda_\mathrm{eff} \, M_\mathrm{Pl}^{-2} +\sqrt{3}-3}{N^2}+O\left(\frac{1}{N^3}\right) \\
& r =\frac{12}{N^2}-\frac{144\, \Lambda_\mathrm{eff} \, M_\mathrm{Pl}^{-2} +12\sqrt{3}}{N^3}+O\left(\frac{1}{N^4}\right)
\end{align}
\end{subequations}
where $N$ is the number of e-folds. Assuming that $\Lambda_\mathrm{eff}$ is the cosmological constant observed today, the combination $\Lambda_\mathrm{eff} \, M_\mathrm{Pl}^{-2}$ is much less than one. Thus the $\Lambda_\mathrm{eff}$-terms do not affect the cosmic fluctuation spectrum and can safely be neglected. The leading contribution to $n_s$ and $r$ are then in agreement with the Planck observational data provided $N\sim50-60$. In particular $n_s\sim0.966$ and $r\sim0.00324$ for $N=60$.

\medskip
\noindent
\emph{NGFPs of Class IIb:} The case of pure gravity with real critical exponents ($\theta_1 \sim 4$ and $\theta_2 \sim 3$) is covered by the Class IIb given in eq.\ \eqref{model2b}. Using the slow-roll approximation and neglecting $\Lambda_\mathrm{eff}$, the spectral index and tensor-to-scalar ratio are approximately given by
\begin{subequations}
\begin{align}
& n_s =1-\frac{2}{N}-\frac{0.85}{N^2}+O\left(\frac{1}{N^{7/3}}\right) \\
& r =\frac{5.33}{N^2}-\frac{6.15}{N^3}+O\left(\frac{1}{N^4}\right)
\end{align}
\end{subequations}
Notably the change in $\theta_2$ and the corresponding eigenvector does not affect the leading $1/N$-term in $n_s$ and appears at order $1/N^2$ only. Setting $N=60$ leads again to values for $(n_s,r)$ compatible with the Planck data: {the value of $n_s$ is identical to the one produced by the Class {IIa} model, $n_s\sim0.966$, }but in this case the tensor-to-scalar ratio is much lower, $r\sim0.00145$ {(see Fig.~\ref{figGUT})}. 

\medskip
\noindent
\emph{NGFPs of Class III:} The grand unified type models are characterized by critical exponents $\theta_i>4$. In this case the leading terms appearing in \eqref{genRGlag} are
\be\label{GUTaction}
S = \frac{M_{\rm Pl}^2}{2} \int d^4x \sqrt{-g} \left(\frac{R^2}{6m^2} + \frac{1}{R^p} \right)
\ee  
where $p=(\theta_2-2)/2\sim2$. 
Although this type of theories do not reproduce General Relativity at long distances, it is interesting to study the corresponding cosmic parameters. 
The values of spectral index and tensor-to-scalar ratio depend on the number of e-folds $N$ and power index $p$. 
Fixing the number of e-folds to $N=60$, we obtained the functions $n_s(p)$ and $r(p)$ (see Fig.~\ref{figGUT}).
As it can be seen in Fig.~\ref{figGUT}, {for $p\gtrsim1$} these models give rise to a power spectrum which is too close to scale-invariance and therefore not compatible with the Planck data. Moreover, the tensor-to-scalar ratio rapidly approaches zero as $p$ increases. 

\medskip
\noindent
\emph{NGFPs of Class I:} The inflationary dynamics resulting from the family of 
effective Lagrangians \eqref{Leffcc} is rather complex. Comparing the coefficients $\tilde b_1, \tilde{b}_2$ and 
$\tilde b_3, \tilde{b}_4, \tilde{b}_5$ one first observes that the latter are suppressed by (inverse) powers of $\xi$. {As the cutoff identification $k^2=\xi R$ does not depend on the particular gravity-matter configuration, we can use the constraint $\xi \sim 10^5$ obtained before. The contribution of $\tilde b_3, \tilde{b}_4, \tilde{b}_5$  can then be neglected  and we set $\tilde b_3 = \tilde{b}_4 = \tilde{b}_5 = 0$.} Fixing $\tilde b_1 = \tilde{b}_2 = 1$ all models {with $\theta^\prime\lesssim3$} lead to an oscillatory scalar potentials supporting an infinite number of meta-stable de Sitter vacua. A sufficiently pronounced inflationary phase leading to a sufficient number of e-folds may then be obtained by placing initial conditions for the scalar field close to a local maximum of the potential. The CMB fluctuations are then created when the field is very close to the local maximum, entailing $r \ll 1$. The value of the spectral index $n_s$ is then essentially determined by the second slow-roll parameter $\eta_V$. 
 Increasing $\theta^\prime$, the oscillations in the potential start smoothing out and the standard nearly-flat plateau is recovered when $\theta^\prime \gtrsim 3$. In particular the values for $(n_s,r)$ are not compatible with observations unless $\theta^\prime \gtrsim 5$. For the specific model \eqref{model1}, the Lagrangian \eqref{Leffcc} gives rise to the scalar potential shown in Fig.\ \ref{fignotGUT}. The slow-roll conditions are fulfilled for all values $\phi \gtrsim 2.19{\,M_\mathrm{Pl}}$. An inflationary phase including an exit from the slow-roll condition can be realized by placing the starting value of $\phi$ close to the maximum at $\phi \simeq 4.23{\,M_\mathrm{Pl}}$. Evaluating the cosmic parameters for an inflationary scenario where the field rolls towards the minimum on the left yields $n_s = 0.423$ and $r \sim 10^{-16}$ which is clearly ruled out by observations. Thus one concludes that while some of the effective Lagrangians \eqref{Leffcc} give rise to a realistic inflationary dynamics, these models are not related to the critical exponents obtained from the renormalization group computations. 

\section{Conclusions}
\label{conclu}
The asymptotic safety mechanism provides a natural way for constructing consistent and predictive high-energy completions of gravity and gravity-matter systems. The characteristic properties of the NGFPs controlling the theories at trans-Planckian energy depend on the specific matter content of the model. Based on the Einstein-Hilbert approximation supplemented by minimally coupled matter fields one can distinguish three different classes of fixed points (see Table \ref{Tab.3}) according to their critical exponents and position. The effective models describing the evolution in the very early universe, eq.\ \eqref{genRGlag}, sensitively depend on this data. The results are quite intuitive and corroborate the picture advocated in \cite{br07}: while the NGFP itself provides a scale-free scalar power spectrum and vanishing scalar-to-tensor ratio, flowing away from the fixed point induces (small) corrections to these values. Studying the resulting NGFP-driven inflationary scenarios and comparing the theoretical predictions with Planck data puts constraints on the fixed point properties. 

{Notably, cosmologies arising from RG-improvements have already undergone detailed investigations at the level of both pure gravity and dilaton-type actions \cite{irfp,br02,dou,guberina03,2004reuterw1,2004reuterw2,
guberina05,resa05,boes,br07,weinberg10,cai11,conpe,alfio12,
AlfioAle1,cai13,2012PhRvD..85d3501C,2016JCAP...02..048S,2017JCAP...07..015T}. The novel feature of this work consists in  a RG-improvement procedure which takes the contribution of (standard model) matter fields into account in a rather systematic way. As a result, the resulting effective cosmological dynamics knows about the particle content of the universe. Quite remarkably, the models whose matter field content is close to the one of the standard model of particle physics are compatible with the Planck data while fixed points associated with GUT-type models do not reproduce the observations. In particular, the discussion around eq.\ \eqref{fpcond1} shows that an acceptable normalization of scalar power spectrum requires $\lambda_* \le 0$ and can not be achieved if $\lambda_* > 0$. }

Remarkably, similar conclusions favoring a negative value of $\lambda_*$ have been reached in \cite{Eichhorn:2017ylw} and \cite{Bonanno:2017gji} albeit based on very different considerations: in the former obtaining phenomenological viable particle physics models based on Asymptotic Safety mechanism required a negative fixed point value of the dimensionless cosmological constant, while in the latter $\lambda_* \le 0$ was the key to a nonsingular cosmological evolution. It would be very interesting to see whether this feature can be corroborated by further studies.


\bigskip
\centerline{\bf Acknowledgments}
\medskip
We thank G.\ Gubitosi for interesting discussions. The research of A.P.\ is supported by the DFG under grant no Ei-1037/1. F.S.\ is supported by the Netherlands Organisation for Scientific Research (NWO) within the Foundation for Fundamental Research on Matter (FOM) grants 13VP12 and 13PR3137.

\bibliography{nk}

\begin{thebibliography}{100}
\expandafter\ifx\csname url\endcsname\relax
  \def\url#1{\texttt{#1}}\fi
\expandafter\ifx\csname urlprefix\endcsname\relax\def\urlprefix{URL }\fi
\expandafter\ifx\csname href\endcsname\relax
  \def\href#1#2{#2} \def\path#1{#1}\fi

\bibitem{pla15:1}
{Planck Collaboration}, P.~A.~R. {Ade}, N.~{Aghanim}, M.~{Arnaud},
  M.~{Ashdown}, J.~{Aumont}, C.~{Baccigalupi}, A.~J. {Banday}, R.~B.
  {Barreiro}, J.~G. {Bartlett}, et~al., {Planck 2015 results. XIII.
  Cosmological parameters}, \aap 594 (2016) A13.
\newblock \href {http://arxiv.org/abs/1502.01589} {\path{arXiv:1502.01589}},
  \href {http://dx.doi.org/10.1051/0004-6361/201525830}
  {\path{doi:10.1051/0004-6361/201525830}}.

\bibitem{pla15:2}
{Planck Collaboration}, P.~A.~R. {Ade}, N.~{Aghanim}, M.~{Arnaud}, F.~{Arroja},
  M.~{Ashdown}, J.~{Aumont}, C.~{Baccigalupi}, M.~{Ballardini}, A.~J. {Banday},
  et~al., {Planck 2015 results. XX. Constraints on inflation}, \aap 594 (2016)
  A20.
\newblock \href {http://arxiv.org/abs/1502.02114} {\path{arXiv:1502.02114}},
  \href {http://dx.doi.org/10.1051/0004-6361/201525898}
  {\path{doi:10.1051/0004-6361/201525898}}.

\bibitem{Patrignani:2016xqp}
C.~Patrignani, et~al., {Review of Particle Physics}, Chin. Phys. C40~(10)
  (2016) 100001.
\newblock \href {http://dx.doi.org/10.1088/1674-1137/40/10/100001}
  {\path{doi:10.1088/1674-1137/40/10/100001}}.

\bibitem{Ade:2015lrj}
P.~A.~R. Ade, et~al., {Planck 2015 results. XX. Constraints on inflation},
  Astron. Astrophys. 594 (2016) A20.
\newblock \href {http://arxiv.org/abs/1502.02114} {\path{arXiv:1502.02114}},
  \href {http://dx.doi.org/10.1051/0004-6361/201525898}
  {\path{doi:10.1051/0004-6361/201525898}}.

\bibitem{inflationaris}
J.~{Martin}, C.~{Ringeval}, V.~{Vennin}, {Encyclop{\ae}dia Inflationaris},
  Physics of the Dark Universe 5 (2014) 75--235.
\newblock \href {http://arxiv.org/abs/1303.3787} {\path{arXiv:1303.3787}},
  \href {http://dx.doi.org/10.1016/j.dark.2014.01.003}
  {\path{doi:10.1016/j.dark.2014.01.003}}.

\bibitem{1976W}
S.~{Weinberg}, {Critical Phenomena for Field Theorists}, Erice Subnuclear
  Physics.

\bibitem{we79}
S.~{Weinberg}, {Ultraviolet divergences in quantum theories of gravitation.},
  in: S.~W. {Hawking}, W.~{Israel} (Eds.), General Relativity: An Einstein
  centenary survey, 1979, pp. 790--831.

\bibitem{Reuter:1996cp}
M.~Reuter, {Nonperturbative evolution equation for quantum gravity}, Phys. Rev.
  D57 (1998) 971--985.
\newblock \href {http://arxiv.org/abs/hep-th/9605030}
  {\path{arXiv:hep-th/9605030}}, \href
  {http://dx.doi.org/10.1103/PhysRevD.57.971}
  {\path{doi:10.1103/PhysRevD.57.971}}.

\bibitem{dou}
D.~{Dou}, R.~{Percacci}, {The running gravitational couplings}, Classical and
  Quantum Gravity 15 (1998) 3449--3468.
\newblock \href {http://arxiv.org/abs/hep-th/9707239}
  {\path{arXiv:hep-th/9707239}}, \href
  {http://dx.doi.org/10.1088/0264-9381/15/11/011}
  {\path{doi:10.1088/0264-9381/15/11/011}}.

\bibitem{so99}
W.~{Souma}, {Non-Trivial Ultraviolet Fixed Point in Quantum Gravity}, Progress
  of Theoretical Physics 102 (1999) 181--195.
\newblock \href {http://arxiv.org/abs/hep-th/9907027}
  {\path{arXiv:hep-th/9907027}}, \href {http://dx.doi.org/10.1143/PTP.102.181}
  {\path{doi:10.1143/PTP.102.181}}.

\bibitem{la02}
O.~{Lauscher}, M.~{Reuter}, {Flow equation of quantum Einstein gravity in a
  higher-derivative truncation}, \prd 66~(2) (2002) 025026.
\newblock \href {http://arxiv.org/abs/hep-th/0205062}
  {\path{arXiv:hep-th/0205062}}, \href
  {http://dx.doi.org/10.1103/PhysRevD.66.025026}
  {\path{doi:10.1103/PhysRevD.66.025026}}.

\bibitem{resa02b}
M.~{Reuter}, F.~{Saueressig}, {A class of nonlocal truncations in quantum
  Einstein gravity and its renormalization group behavior}, \prd 66~(12) (2002)
  125001.
\newblock \href {http://arxiv.org/abs/hep-th/0206145}
  {\path{arXiv:hep-th/0206145}}, \href
  {http://dx.doi.org/10.1103/PhysRevD.66.125001}
  {\path{doi:10.1103/PhysRevD.66.125001}}.

\bibitem{resa02}
M.~{Reuter}, F.~{Saueressig}, {Renormalization group flow of quantum gravity in
  the Einstein-Hilbert truncation}, \prd 65~(6) (2002) 065016.
\newblock \href {http://arxiv.org/abs/hep-th/0110054}
  {\path{arXiv:hep-th/0110054}}, \href
  {http://dx.doi.org/10.1103/PhysRevD.65.065016}
  {\path{doi:10.1103/PhysRevD.65.065016}}.

\bibitem{daniel04}
D.~F. {Litim}, {Fixed Points of Quantum Gravity}, Physical Review Letters
  92~(20) (2004) 201301.
\newblock \href {http://arxiv.org/abs/hep-th/0312114}
  {\path{arXiv:hep-th/0312114}}, \href
  {http://dx.doi.org/10.1103/PhysRevLett.92.201301}
  {\path{doi:10.1103/PhysRevLett.92.201301}}.

\bibitem{bore05}
A.~{Bonanno}, M.~{Reuter}, {Proper time flow equation for gravity}, Journal of
  High Energy Physics 2 (2005) 35.
\newblock \href {http://arxiv.org/abs/hep-th/0410191}
  {\path{arXiv:hep-th/0410191}}, \href
  {http://dx.doi.org/10.1088/1126-6708/2005/02/035}
  {\path{doi:10.1088/1126-6708/2005/02/035}}.

\bibitem{co09}
A.~{Codello}, R.~{Percacci}, C.~{Rahmede}, {Investigating the ultraviolet
  properties of gravity with a Wilsonian renormalization group equation},
  Annals of Physics 324 (2009) 414--469.
\newblock \href {http://arxiv.org/abs/0805.2909} {\path{arXiv:0805.2909}},
  \href {http://dx.doi.org/10.1016/j.aop.2008.08.008}
  {\path{doi:10.1016/j.aop.2008.08.008}}.

\bibitem{bema10}
D.~{Benedetti}, P.~F. {Machado}, F.~{Saueressig}, {Taming perturbative
  divergences in asymptotically safe gravity}, Nuclear Physics B 824 (2010)
  168--191.
\newblock \href {http://arxiv.org/abs/0902.4630} {\path{arXiv:0902.4630}},
  \href {http://dx.doi.org/10.1016/j.nuclphysb.2009.08.023}
  {\path{doi:10.1016/j.nuclphysb.2009.08.023}}.

\bibitem{mare11}
E.~{Manrique}, M.~{Reuter}, F.~{Saueressig}, {Matter induced bimetric actions
  for gravity}, Annals of Physics 326 (2011) 440--462.
\newblock \href {http://arxiv.org/abs/1003.5129} {\path{arXiv:1003.5129}},
  \href {http://dx.doi.org/10.1016/j.aop.2010.11.003}
  {\path{doi:10.1016/j.aop.2010.11.003}}.

\bibitem{ast13}
A.~{Eichhorn}, {Faddeev-Popov ghosts in quantum gravity beyond perturbation
  theory}, \prd 87~(12) (2013) 124016.
\newblock \href {http://arxiv.org/abs/1301.0632} {\path{arXiv:1301.0632}},
  \href {http://dx.doi.org/10.1103/PhysRevD.87.124016}
  {\path{doi:10.1103/PhysRevD.87.124016}}.

\bibitem{beca}
D.~{Benedetti}, F.~{Caravelli}, {The local potential approximation in quantum
  gravity}, Journal of High Energy Physics 6 (2012) 17.
\newblock \href {http://arxiv.org/abs/1204.3541} {\path{arXiv:1204.3541}},
  \href {http://dx.doi.org/10.1007/JHEP06(2012)017}
  {\path{doi:10.1007/JHEP06(2012)017}}.

\bibitem{bogu}
A.~{Bonanno}, F.~{Guarnieri}, {Universality and symmetry breaking in
  conformally reduced quantum gravity}, \prd 86~(10) (2012) 105027.
\newblock \href {http://arxiv.org/abs/1206.6531} {\path{arXiv:1206.6531}},
  \href {http://dx.doi.org/10.1103/PhysRevD.86.105027}
  {\path{doi:10.1103/PhysRevD.86.105027}}.

\bibitem{begu14}
D.~{Benedetti}, F.~{Guarnieri}, {One-loop renormalization in a toy model of
  Ho{\v r}ava-Lifshitz gravity}, Journal of High Energy Physics 3 (2014) 78.
\newblock \href {http://arxiv.org/abs/1311.6253} {\path{arXiv:1311.6253}},
  \href {http://dx.doi.org/10.1007/JHEP03(2014)078}
  {\path{doi:10.1007/JHEP03(2014)078}}.

\bibitem{begu}
D.~{Benedetti}, F.~{Guarnieri}, {Brans-Dicke theory in the local potential
  approximation}, New Journal of Physics 16~(5) (2014) 053051.
\newblock \href {http://arxiv.org/abs/1311.1081} {\path{arXiv:1311.1081}},
  \href {http://dx.doi.org/10.1088/1367-2630/16/5/053051}
  {\path{doi:10.1088/1367-2630/16/5/053051}}.

\bibitem{demme}
M.~{Demmel}, F.~{Saueressig}, O.~{Zanusso}, {RG flows of Quantum Einstein
  Gravity on maximally symmetric spaces}, Journal of High Energy Physics 6
  (2014) 26.
\newblock \href {http://arxiv.org/abs/1401.5495} {\path{arXiv:1401.5495}},
  \href {http://dx.doi.org/10.1007/JHEP06(2014)026}
  {\path{doi:10.1007/JHEP06(2014)026}}.

\bibitem{dimo15}
J.~A. {Dietz}, T.~R. {Morris}, {Background independent exact renormalization
  group for conformally reduced gravity}, Journal of High Energy Physics 4
  (2015) 118.
\newblock \href {http://arxiv.org/abs/1502.07396} {\path{arXiv:1502.07396}},
  \href {http://dx.doi.org/10.1007/JHEP04(2015)118}
  {\path{doi:10.1007/JHEP04(2015)118}}.

\bibitem{Souma:1999at}
W.~{Souma}, {Non-Trivial Ultraviolet Fixed Point in Quantum Gravity}, Progress
  of Theoretical Physics 102 (1999) 181--195.
\newblock \href {http://arxiv.org/abs/hep-th/9907027}
  {\path{arXiv:hep-th/9907027}}, \href {http://dx.doi.org/10.1143/PTP.102.181}
  {\path{doi:10.1143/PTP.102.181}}.

\bibitem{Codello:2007bd}
A.~{Codello}, R.~{Percacci}, C.~{Rahmede}, {Ultraviolet Properties of
  f(R)-gravity}, International Journal of Modern Physics A 23 (2008) 143--150.
\newblock \href {http://arxiv.org/abs/0705.1769} {\path{arXiv:0705.1769}},
  \href {http://dx.doi.org/10.1142/S0217751X08038135}
  {\path{doi:10.1142/S0217751X08038135}}.

\bibitem{Machado:2007ea}
P.~F. Machado, F.~Saueressig, {On the renormalization group flow of
  f(R)-gravity}, \prd 77 (2008) 124045.
\newblock \href {http://arxiv.org/abs/arXiv:0712.0445}
  {\path{arXiv:arXiv:0712.0445}}, \href
  {http://dx.doi.org/10.1103/PhysRevD.77.124045}
  {\path{doi:10.1103/PhysRevD.77.124045}}.

\bibitem{Falls:2013bv}
K.~{Falls}, D.~F. {Litim}, K.~{Nikolakopoulos}, C.~{Rahmede}, {A bootstrap
  strategy for asymptotic safety}\href {http://arxiv.org/abs/1301.4191}
  {\path{arXiv:1301.4191}}.

\bibitem{ast15}
A.~{Eichhorn}, {The Renormalization Group flow of unimodular f(R) gravity},
  Journal of High Energy Physics 4 (2015) 96.
\newblock \href {http://arxiv.org/abs/1501.05848} {\path{arXiv:1501.05848}},
  \href {http://dx.doi.org/10.1007/JHEP04(2015)096}
  {\path{doi:10.1007/JHEP04(2015)096}}.

\bibitem{Falls:2016msz}
K.~{Falls}, N.~{Ohta}, {Renormalization group equation for f(R) gravity on
  hyperbolic spaces}, \prd 94~(8) (2016) 084005.
\newblock \href {http://arxiv.org/abs/1607.08460} {\path{arXiv:1607.08460}},
  \href {http://dx.doi.org/10.1103/PhysRevD.94.084005}
  {\path{doi:10.1103/PhysRevD.94.084005}}.

\bibitem{Dietz:2012ic}
J.~A. {Dietz}, T.~R. {Morris}, {Asymptotic safety in the f(R) approximation},
  Journal of High Energy Physics 1 (2013) 108.
\newblock \href {http://arxiv.org/abs/1211.0955} {\path{arXiv:1211.0955}},
  \href {http://dx.doi.org/10.1007/JHEP01(2013)108}
  {\path{doi:10.1007/JHEP01(2013)108}}.

\bibitem{Ohta:2015efa}
N.~{Ohta}, R.~{Percacci}, G.~P. {Vacca}, {Flow equation for f(R) gravity and
  some of its exact solutions}, \prd 92~(6) (2015) 061501.
\newblock \href {http://arxiv.org/abs/1507.00968} {\path{arXiv:1507.00968}},
  \href {http://dx.doi.org/10.1103/PhysRevD.92.061501}
  {\path{doi:10.1103/PhysRevD.92.061501}}.

\bibitem{Ohta:2015fcu}
N.~{Ohta}, R.~{Percacci}, G.~P. {Vacca}, {Renormalization group equation and
  scaling solutions for f( R) gravity in exponential parametrization}, European
  Physical Journal C 76 (2016) 46.
\newblock \href {http://arxiv.org/abs/1511.09393} {\path{arXiv:1511.09393}},
  \href {http://dx.doi.org/10.1140/epjc/s10052-016-3895-1}
  {\path{doi:10.1140/epjc/s10052-016-3895-1}}.

\bibitem{Codello:2006in}
A.~{Codello}, R.~{Percacci}, {Fixed Points of Higher-Derivative Gravity},
  Physical Review Letters 97~(22) (2006) 221301.
\newblock \href {http://arxiv.org/abs/hep-th/0607128}
  {\path{arXiv:hep-th/0607128}}, \href
  {http://dx.doi.org/10.1103/PhysRevLett.97.221301}
  {\path{doi:10.1103/PhysRevLett.97.221301}}.

\bibitem{Benedetti:2009rx}
D.~{Benedetti}, P.~F. {Machado}, F.~{Saueressig}, {Asymptotic Safety in
  Higher-Derivative Gravity}, Modern Physics Letters A 24 (2009) 2233--2241.
\newblock \href {http://arxiv.org/abs/0901.2984} {\path{arXiv:0901.2984}},
  \href {http://dx.doi.org/10.1142/S0217732309031521}
  {\path{doi:10.1142/S0217732309031521}}.

\bibitem{Groh:2011vn}
F.~{Saueressig}, K.~{Groh}, S.~{Rechenberger}, O.~{Zanusso}, {Higher Derivative
  Gravity from the Universal Renormalization Group Machine}, ArXiv
  e-prints\href {http://arxiv.org/abs/1111.1743} {\path{arXiv:1111.1743}}.

\bibitem{Gies:2016con}
H.~{Gies}, B.~{Knorr}, S.~{Lippoldt}, F.~{Saueressig}, {Gravitational Two-Loop
  Counterterm Is Asymptotically Safe}, Physical Review Letters 116~(21) (2016)
  211302.
\newblock \href {http://arxiv.org/abs/1601.01800} {\path{arXiv:1601.01800}},
  \href {http://dx.doi.org/10.1103/PhysRevLett.116.211302}
  {\path{doi:10.1103/PhysRevLett.116.211302}}.

\bibitem{Rechenberger:2012pm}
S.~{Rechenberger}, F.~{Saueressig}, {$R^2$ phase diagram of quantum Einstein
  gravity and its spectral dimension}, \prd 86~(2) (2012) 024018.
\newblock \href {http://arxiv.org/abs/1206.0657} {\path{arXiv:1206.0657}},
  \href {http://dx.doi.org/10.1103/PhysRevD.86.024018}
  {\path{doi:10.1103/PhysRevD.86.024018}}.

\bibitem{Christiansen:2012rx}
N.~{Christiansen}, D.~F. {Litim}, J.~M. {Pawlowski}, A.~{Rodigast}, {Fixed
  points and infrared completion of quantum gravity}, Physics Letters B 728
  (2014) 114--117.
\newblock \href {http://dx.doi.org/10.1016/j.physletb.2013.11.025}
  {\path{doi:10.1016/j.physletb.2013.11.025}}.

\bibitem{Manrique:2011jc}
E.~{Manrique}, S.~{Rechenberger}, F.~{Saueressig}, {Asymptotically Safe
  Lorentzian Gravity}, Physical Review Letters 106~(25) (2011) 251302.
\newblock \href {http://arxiv.org/abs/1102.5012} {\path{arXiv:1102.5012}},
  \href {http://dx.doi.org/10.1103/PhysRevLett.106.251302}
  {\path{doi:10.1103/PhysRevLett.106.251302}}.

\bibitem{AleFrank1}
J.~{Biemans}, A.~{Platania}, F.~{Saueressig}, {Quantum gravity on foliated
  spacetimes: Asymptotically safe and sound}, \prd 95~(8) (2017) 086013.
\newblock \href {http://arxiv.org/abs/1609.04813} {\path{arXiv:1609.04813}},
  \href {http://dx.doi.org/10.1103/PhysRevD.95.086013}
  {\path{doi:10.1103/PhysRevD.95.086013}}.

\bibitem{AleFrank2}
J.~{Biemans}, A.~{Platania}, F.~{Saueressig}, {Renormalization group fixed
  points of foliated gravity-matter systems}, Journal of High Energy Physics 5
  (2017) 93.
\newblock \href {http://arxiv.org/abs/1702.06539} {\path{arXiv:1702.06539}},
  \href {http://dx.doi.org/10.1007/JHEP05(2017)093}
  {\path{doi:10.1007/JHEP05(2017)093}}.

\bibitem{Platania:2017djo}
A.~Platania, F.~Saueressig, {Functional Renormalization Group flows on
  Friedman-Lema\^{\i}tre-Robertson-Walker backgrounds}\href
  {http://arxiv.org/abs/1710.01972} {\path{arXiv:1710.01972}}.

\bibitem{2017franktopos}
W.~B. {Houthoff}, A.~{Kurov}, F.~{Saueressig}, {Impact of topology in foliated
  quantum Einstein gravity}, European Physical Journal C 77 (2017) 491.
\newblock \href {http://arxiv.org/abs/1705.01848} {\path{arXiv:1705.01848}},
  \href {http://dx.doi.org/10.1140/epjc/s10052-017-5046-8}
  {\path{doi:10.1140/epjc/s10052-017-5046-8}}.

\bibitem{Percacci:2002ie}
R.~{Percacci}, D.~{Perini}, {Constraints on matter from asymptotic safety},
  \prd 67~(8) (2003) 081503.
\newblock \href {http://arxiv.org/abs/hep-th/0207033}
  {\path{arXiv:hep-th/0207033}}, \href
  {http://dx.doi.org/10.1103/PhysRevD.67.081503}
  {\path{doi:10.1103/PhysRevD.67.081503}}.

\bibitem{Percacci:2003jz}
R.~{Percacci}, D.~{Perini}, {Asymptotic safety of gravity coupled to matter},
  \prd 68~(4) (2003) 044018.
\newblock \href {http://arxiv.org/abs/hep-th/0304222}
  {\path{arXiv:hep-th/0304222}}, \href
  {http://dx.doi.org/10.1103/PhysRevD.68.044018}
  {\path{doi:10.1103/PhysRevD.68.044018}}.

\bibitem{Zanusso:2009bs}
O.~{Zanusso}, L.~{Zambelli}, G.~P. {Vacca}, R.~{Percacci}, {Gravitational
  corrections to Yukawa systems}, Physics Letters B 689 (2010) 90--94.
\newblock \href {http://arxiv.org/abs/0904.0938} {\path{arXiv:0904.0938}},
  \href {http://dx.doi.org/10.1016/j.physletb.2010.04.043}
  {\path{doi:10.1016/j.physletb.2010.04.043}}.

\bibitem{Vacca:2010mj}
G.~P. {Vacca}, O.~{Zanusso}, {Asymptotic Safety in Einstein Gravity and
  Scalar-Fermion Matter}, Physical Review Letters 105~(23) (2010) 231601.
\newblock \href {http://arxiv.org/abs/1009.1735} {\path{arXiv:1009.1735}},
  \href {http://dx.doi.org/10.1103/PhysRevLett.105.231601}
  {\path{doi:10.1103/PhysRevLett.105.231601}}.

\bibitem{Harst:2011zx}
U.~{Harst}, M.~{Reuter}, {QED coupled to QEG}, Journal of High Energy Physics 5
  (2011) 119.
\newblock \href {http://arxiv.org/abs/1101.6007} {\path{arXiv:1101.6007}},
  \href {http://dx.doi.org/10.1007/JHEP05(2011)119}
  {\path{doi:10.1007/JHEP05(2011)119}}.

\bibitem{Eichhorn:2011pc}
A.~{Eichhorn}, H.~{Gies}, {Light fermions in quantum gravity}, New Journal of
  Physics 13~(12) (2011) 125012.
\newblock \href {http://arxiv.org/abs/1104.5366} {\path{arXiv:1104.5366}},
  \href {http://dx.doi.org/10.1088/1367-2630/13/12/125012}
  {\path{doi:10.1088/1367-2630/13/12/125012}}.

\bibitem{Dona:2012am}
P.~{Don{\`a}}, R.~{Percacci}, {Functional renormalization with fermions and
  tetrads}, \prd 87~(4) (2013) 045002.
\newblock \href {http://arxiv.org/abs/1209.3649} {\path{arXiv:1209.3649}},
  \href {http://dx.doi.org/10.1103/PhysRevD.87.045002}
  {\path{doi:10.1103/PhysRevD.87.045002}}.

\bibitem{Dona:2013qba}
P.~{Don{\`a}}, A.~{Eichhorn}, R.~{Percacci}, {Matter matters in asymptotically
  safe quantum gravity}, \prd 89~(8) (2014) 084035.
\newblock \href {http://arxiv.org/abs/1311.2898} {\path{arXiv:1311.2898}},
  \href {http://dx.doi.org/10.1103/PhysRevD.89.084035}
  {\path{doi:10.1103/PhysRevD.89.084035}}.

\bibitem{Dona:2014pla}
P.~{Don{\`a}}, A.~{Eichhorn}, R.~{Percacci}, {Consistency of matter models with
  asymptotically safe quantum gravity}, Canadian Journal of Physics 93 (2015)
  988--994.
\newblock \href {http://arxiv.org/abs/1410.4411} {\path{arXiv:1410.4411}},
  \href {http://dx.doi.org/10.1139/cjp-2014-0574}
  {\path{doi:10.1139/cjp-2014-0574}}.

\bibitem{peva14}
R.~{Percacci}, G.~P. {Vacca}, {Are there scaling solutions in the O(N)-models
  for large N in d>4 ?}, \prd 90~(10) (2014) 107702.
\newblock \href {http://arxiv.org/abs/1405.6622} {\path{arXiv:1405.6622}},
  \href {http://dx.doi.org/10.1103/PhysRevD.90.107702}
  {\path{doi:10.1103/PhysRevD.90.107702}}.

\bibitem{Meibohm:2015twa}
J.~{Meibohm}, J.~M. {Pawlowski}, M.~{Reichert}, {Asymptotic safety of
  gravity-matter systems}, \prd 93~(8) (2016) 084035.
\newblock \href {http://arxiv.org/abs/1510.07018} {\path{arXiv:1510.07018}},
  \href {http://dx.doi.org/10.1103/PhysRevD.93.084035}
  {\path{doi:10.1103/PhysRevD.93.084035}}.

\bibitem{Oda:2015sma}
K.~Y. {Oda}, M.~{Yamada}, {Non-minimal coupling in Higgs-Yukawa model with
  asymptotically safe gravity}, Classical and Quantum Gravity 33~(12) (2016)
  125011.
\newblock \href {http://arxiv.org/abs/1510.03734} {\path{arXiv:1510.03734}},
  \href {http://dx.doi.org/10.1088/0264-9381/33/12/125011}
  {\path{doi:10.1088/0264-9381/33/12/125011}}.

\bibitem{Dona:2015tnf}
P.~{Don{\`a}}, A.~{Eichhorn}, P.~{Labus}, R.~{Percacci}, {Asymptotic safety in
  an interacting system of gravity and scalar matter}, \prd 93~(4) (2016)
  044049.
\newblock \href {http://arxiv.org/abs/1512.01589} {\path{arXiv:1512.01589}},
  \href {http://dx.doi.org/10.1103/PhysRevD.93.044049}
  {\path{doi:10.1103/PhysRevD.93.044049}}.

\bibitem{Labus:2015ska}
P.~{Labus}, R.~{Percacci}, G.~P. {Vacca}, {Asymptotic safety in O(N) scalar
  models coupled to gravity}, Physics Letters B 753 (2016) 274--281.
\newblock \href {http://arxiv.org/abs/1505.05393} {\path{arXiv:1505.05393}},
  \href {http://dx.doi.org/10.1016/j.physletb.2015.12.022}
  {\path{doi:10.1016/j.physletb.2015.12.022}}.

\bibitem{Meibohm:2016mkp}
J.~{Meibohm}, J.~M. {Pawlowski}, {Chiral fermions in asymptotically safe
  quantum gravity}, European Physical Journal C 76 (2016) 285.
\newblock \href {http://arxiv.org/abs/1601.04597} {\path{arXiv:1601.04597}},
  \href {http://dx.doi.org/10.1140/epjc/s10052-016-4132-7}
  {\path{doi:10.1140/epjc/s10052-016-4132-7}}.

\bibitem{Eichhorn:2016esv}
A.~{Eichhorn}, A.~{Held}, J.~M. {Pawlowski}, {Quantum-gravity effects on a
  Higgs-Yukawa model}, \prd 94~(10) (2016) 104027.
\newblock \href {http://arxiv.org/abs/1604.02041} {\path{arXiv:1604.02041}},
  \href {http://dx.doi.org/10.1103/PhysRevD.94.104027}
  {\path{doi:10.1103/PhysRevD.94.104027}}.

\bibitem{Eichhorn:2017ylw}
A.~Eichhorn, A.~Held, {Top mass from asymptotic safety}\href
  {http://arxiv.org/abs/1707.01107} {\path{arXiv:1707.01107}}.

\bibitem{Alkofer:2018fxj}
N.~Alkofer, F.~Saueressig, {Asymptotically safe $f(R)$-gravity coupled to
  matter I: the polynomial case}\href {http://arxiv.org/abs/1802.00498}
  {\path{arXiv:1802.00498}}.

\bibitem{AlfioAle1}
A.~{Bonanno}, A.~{Platania}, {Asymptotically safe inflation from quadratic
  gravity}, Physics Letters B 750 (2015) 638--642.
\newblock \href {http://arxiv.org/abs/1507.03375} {\path{arXiv:1507.03375}},
  \href {http://dx.doi.org/10.1016/j.physletb.2015.10.005}
  {\path{doi:10.1016/j.physletb.2015.10.005}}.

\bibitem{irfp}
A.~{Bonanno}, M.~{Reuter}, {Cosmology with self-adjusting vacuum energy density
  from a renormalization group fixed point}, Physics Letters B 527 (2002)
  9--17.
\newblock \href {http://arxiv.org/abs/astro-ph/0106468}
  {\path{arXiv:astro-ph/0106468}}, \href
  {http://dx.doi.org/10.1016/S0370-2693(01)01522-2}
  {\path{doi:10.1016/S0370-2693(01)01522-2}}.

\bibitem{br02}
A.~{Bonanno}, M.~{Reuter}, {Cosmology of the Planck era from a renormalization
  group for quantum gravity}, \prd 65~(4) (2002) 043508.
\newblock \href {http://arxiv.org/abs/hep-th/0106133}
  {\path{arXiv:hep-th/0106133}}, \href
  {http://dx.doi.org/10.1103/PhysRevD.65.043508}
  {\path{doi:10.1103/PhysRevD.65.043508}}.

\bibitem{guberina03}
B.~{Guberina}, R.~{Horvat}, H.~{{\v S}tefan{\v c}i{\'c}},
  {Renormalization-group running of the cosmological constant and the fate of
  the universe}, \prd 67~(8) (2003) 083001.
\newblock \href {http://arxiv.org/abs/hep-ph/0211184}
  {\path{arXiv:hep-ph/0211184}}, \href
  {http://dx.doi.org/10.1103/PhysRevD.67.083001}
  {\path{doi:10.1103/PhysRevD.67.083001}}.

\bibitem{2004reuterw1}
M.~{Reuter}, H.~{Weyer}, {Running Newton constant, improved gravitational
  actions, and galaxy rotation curves}, \prd 70~(12) (2004) 124028.
\newblock \href {http://arxiv.org/abs/hep-th/0410117}
  {\path{arXiv:hep-th/0410117}}, \href
  {http://dx.doi.org/10.1103/PhysRevD.70.124028}
  {\path{doi:10.1103/PhysRevD.70.124028}}.

\bibitem{2004reuterw2}
M.~{Reuter}, H.~{Weyer}, {Renormalization group improved gravitational actions:
  A Brans-Dicke approach}, \prd 69~(10) (2004) 104022.
\newblock \href {http://arxiv.org/abs/hep-th/0311196}
  {\path{arXiv:hep-th/0311196}}, \href
  {http://dx.doi.org/10.1103/PhysRevD.69.104022}
  {\path{doi:10.1103/PhysRevD.69.104022}}.

\bibitem{guberina05}
A.~{Babi{\'c}}, B.~{Guberina}, R.~{Horvat}, H.~{{\v S}tefan{\v c}i{\'c}},
  {Renormalization-group running cosmologies: A scale-setting procedure}, \prd
  71~(12) (2005) 124041.
\newblock \href {http://arxiv.org/abs/astro-ph/0407572}
  {\path{arXiv:astro-ph/0407572}}, \href
  {http://dx.doi.org/10.1103/PhysRevD.71.124041}
  {\path{doi:10.1103/PhysRevD.71.124041}}.

\bibitem{resa05}
M.~{Reuter}, F.~{Saueressig}, {From big bang to asymptotic de Sitter: complete
  cosmologies in a quantum gravity framework}, \jcap 9 (2005) 12.
\newblock \href {http://arxiv.org/abs/hep-th/0507167}
  {\path{arXiv:hep-th/0507167}}, \href
  {http://dx.doi.org/10.1088/1475-7516/2005/09/012}
  {\path{doi:10.1088/1475-7516/2005/09/012}}.

\bibitem{boes}
A.~{Bonanno}, G.~{Esposito}, C.~{Rubano}, P.~{Scudellaro}, {The accelerated
  expansion of the universe as a crossover phenomenon}, Classical and Quantum
  Gravity 23 (2006) 3103--3110.
\newblock \href {http://arxiv.org/abs/astro-ph/0507670}
  {\path{arXiv:astro-ph/0507670}}, \href
  {http://dx.doi.org/10.1088/0264-9381/23/9/020}
  {\path{doi:10.1088/0264-9381/23/9/020}}.

\bibitem{br07}
A.~{Bonanno}, M.~{Reuter}, {Entropy signature of the running cosmological
  constant}, JCAP 8 (2007) 024.
\newblock \href {http://arxiv.org/abs/0706.0174} {\path{arXiv:0706.0174}},
  \href {http://dx.doi.org/10.1088/1475-7516/2007/08/024}
  {\path{doi:10.1088/1475-7516/2007/08/024}}.

\bibitem{weinberg10}
S.~{Weinberg}, {Asymptotically safe inflation}, \prd 81~(8) (2010) 083535.
\newblock \href {http://arxiv.org/abs/0911.3165} {\path{arXiv:0911.3165}},
  \href {http://dx.doi.org/10.1103/PhysRevD.81.083535}
  {\path{doi:10.1103/PhysRevD.81.083535}}.

\bibitem{cai11}
Y.~F. {Cai}, D.~A. {Easson}, {Asymptotically safe gravity as a scalar-tensor
  theory and its cosmological implications}, \prd 84~(10) (2011) 103502.
\newblock \href {http://arxiv.org/abs/1107.5815} {\path{arXiv:1107.5815}},
  \href {http://dx.doi.org/10.1103/PhysRevD.84.103502}
  {\path{doi:10.1103/PhysRevD.84.103502}}.

\bibitem{conpe}
A.~{Bonanno}, A.~{Contillo}, R.~{Percacci}, {Inflationary solutions in
  asymptotically safe f(R) theories}, Classical and Quantum Gravity 28~(14)
  (2011) 145026.
\newblock \href {http://arxiv.org/abs/1006.0192} {\path{arXiv:1006.0192}},
  \href {http://dx.doi.org/10.1088/0264-9381/28/14/145026}
  {\path{doi:10.1088/0264-9381/28/14/145026}}.

\bibitem{alfio12}
A.~{Bonanno}, {An effective action for asymptotically safe gravity}, \prd
  85~(8) (2012) 081503.
\newblock \href {http://arxiv.org/abs/1203.1962} {\path{arXiv:1203.1962}},
  \href {http://dx.doi.org/10.1103/PhysRevD.85.081503}
  {\path{doi:10.1103/PhysRevD.85.081503}}.

\bibitem{cai13}
Y.~F. {Cai}, Y.~C. {Chang}, P.~{Chen}, D.~A. {Easson}, T.~{Qiu}, {Planck
  constraints on Higgs modulated reheating of renormalization group improved
  inflation}, \prd 88~(8) (2013) 083508.
\newblock \href {http://arxiv.org/abs/1304.6938} {\path{arXiv:1304.6938}},
  \href {http://dx.doi.org/10.1103/PhysRevD.88.083508}
  {\path{doi:10.1103/PhysRevD.88.083508}}.

\bibitem{2012PhRvD..85d3501C}
A.~{Contillo}, M.~{Hindmarsh}, C.~{Rahmede}, {Renormalization group improvement
  of scalar field inflation}, \prd 85~(4) (2012) 043501.
\newblock \href {http://arxiv.org/abs/1108.0422} {\path{arXiv:1108.0422}},
  \href {http://dx.doi.org/10.1103/PhysRevD.85.043501}
  {\path{doi:10.1103/PhysRevD.85.043501}}.

\bibitem{2016JCAP...02..048S}
I.~D. {Saltas}, {Higgs inflation and quantum gravity: an exact renormalisation
  group approach}, \jcap 2 (2016) 048.
\newblock \href {http://arxiv.org/abs/1512.06134} {\path{arXiv:1512.06134}},
  \href {http://dx.doi.org/10.1088/1475-7516/2016/02/048}
  {\path{doi:10.1088/1475-7516/2016/02/048}}.

\bibitem{2017JCAP...07..015T}
A.~{Tronconi}, {Asymptotically safe non-minimal inflation}, \jcap 7 (2017) 015.
\newblock \href {http://arxiv.org/abs/1704.05312} {\path{arXiv:1704.05312}},
  \href {http://dx.doi.org/10.1088/1475-7516/2017/07/015}
  {\path{doi:10.1088/1475-7516/2017/07/015}}.

\bibitem{2017alfr}
A.~{Bonanno}, F.~{Saueressig}, {Asymptotically safe cosmology - A status
  report}, Comptes Rendus Physique 18 (2017) 254--264.
\newblock \href {http://arxiv.org/abs/1702.04137} {\path{arXiv:1702.04137}},
  \href {http://dx.doi.org/10.1016/j.crhy.2017.02.002}
  {\path{doi:10.1016/j.crhy.2017.02.002}}.

\bibitem{Arnowitt:1962hi}
R.~L. Arnowitt, S.~Deser, C.~W. Misner, {The Dynamics of general relativity},
  Gen. Rel. Grav. 40 (2008) 1997--2027.
\newblock \href {http://arxiv.org/abs/gr-qc/0405109}
  {\path{arXiv:gr-qc/0405109}}, \href
  {http://dx.doi.org/10.1007/s10714-008-0661-1}
  {\path{doi:10.1007/s10714-008-0661-1}}.

\bibitem{Christiansen:2014raa}
N.~Christiansen, B.~Knorr, J.~M. Pawlowski, A.~Rodigast, {Global Flows in
  Quantum Gravity}, Phys. Rev. D93~(4) (2016) 044036.
\newblock \href {http://arxiv.org/abs/1403.1232} {\path{arXiv:1403.1232}},
  \href {http://dx.doi.org/10.1103/PhysRevD.93.044036}
  {\path{doi:10.1103/PhysRevD.93.044036}}.

\bibitem{fayos11}
F.~{Fayos}, R.~{Torres}, {A quantum improvement to the gravitational collapse
  of radiating stars}, Classical and Quantum Gravity 28~(10) (2011) 105004.
\newblock \href {http://dx.doi.org/10.1088/0264-9381/28/10/105004}
  {\path{doi:10.1088/0264-9381/28/10/105004}}.

\bibitem{br00}
A.~{Bonanno}, M.~{Reuter}, {Renormalization group improved black hole
  spacetimes}, \prd 62~(4) (2000) 043008.
\newblock \href {http://arxiv.org/abs/hep-th/0002196}
  {\path{arXiv:hep-th/0002196}}, \href
  {http://dx.doi.org/10.1103/PhysRevD.62.043008}
  {\path{doi:10.1103/PhysRevD.62.043008}}.

\bibitem{Reuter:2004nx}
M.~{Reuter}, H.~{Weyer}, {Quantum gravity at astrophysical distances?}, \jcap
  12 (2004) 001.
\newblock \href {http://arxiv.org/abs/hep-th/0410119}
  {\path{arXiv:hep-th/0410119}}, \href
  {http://dx.doi.org/10.1088/1475-7516/2004/12/001}
  {\path{doi:10.1088/1475-7516/2004/12/001}}.

\bibitem{cw}
S.~{Coleman}, E.~{Weinberg}, {Radiative Corrections as the Origin of
  Spontaneous Symmetry Breaking}, \prd 7 (1973) 1888--1910.
\newblock \href {http://dx.doi.org/10.1103/PhysRevD.7.1888}
  {\path{doi:10.1103/PhysRevD.7.1888}}.

\bibitem{1973migdal}
A.~B. {Migdal}, {Vacuum polarization in strong non-homogeneous fields}, Nuclear
  Physics B 52 (1973) 483--505.
\newblock \href {http://dx.doi.org/10.1016/0550-3213(73)90575-0}
  {\path{doi:10.1016/0550-3213(73)90575-0}}.

\bibitem{1973gross}
D.~J. {Gross}, F.~{Wilczek}, {Asymptotically Free Gauge Theories. I}, \prd 8
  (1973) 3633--3652.
\newblock \href {http://dx.doi.org/10.1103/PhysRevD.8.3633}
  {\path{doi:10.1103/PhysRevD.8.3633}}.

\bibitem{1978pagels}
H.~{Pagels}, E.~{Tomboulis}, {Vacuum of the quantum Yang-Mills theory and
  magnetostatics}, Nuclear Physics B 143 (1978) 485--502.
\newblock \href {http://dx.doi.org/10.1016/0550-3213(78)90065-2}
  {\path{doi:10.1016/0550-3213(78)90065-2}}.

\bibitem{1978Matinyan}
S.~G. {Matinyan}, G.~K. {Savvidy}, {Vacuum polarization induced by the intense
  gauge field}, Nuclear Physics B 134 (1978) 539--545.
\newblock \href {http://dx.doi.org/10.1016/0550-3213(78)90463-7}
  {\path{doi:10.1016/0550-3213(78)90463-7}}.

\bibitem{1983adler}
S.~L. {Adler}, {Short-distance perturbation theory for the leading logarithm
  models}, Nuclear Physics B 217 (1983) 381--394.
\newblock \href {http://dx.doi.org/10.1016/0550-3213(83)90153-0}
  {\path{doi:10.1016/0550-3213(83)90153-0}}.

\bibitem{saltas12}
M.~{Hindmarsh}, I.~D. {Saltas}, {f(R) gravity from the renormalization group},
  \prd 86~(6) (2012) 064029.
\newblock \href {http://arxiv.org/abs/1203.3957} {\path{arXiv:1203.3957}},
  \href {http://dx.doi.org/10.1103/PhysRevD.86.064029}
  {\path{doi:10.1103/PhysRevD.86.064029}}.

\bibitem{co15}
E.~J. {Copeland}, C.~{Rahmede}, I.~D. {Saltas}, {Asymptotically safe
  Starobinsky inflation}, \prd 91~(10) (2015) 103530.
\newblock \href {http://arxiv.org/abs/1311.0881} {\path{arXiv:1311.0881}},
  \href {http://dx.doi.org/10.1103/PhysRevD.91.103530}
  {\path{doi:10.1103/PhysRevD.91.103530}}.

\bibitem{Koch:2014cqa}
B.~Koch, F.~Saueressig, {Black holes within Asymptotic Safety}, Int. J. Mod.
  Phys. A29~(8) (2014) 1430011.
\newblock \href {http://arxiv.org/abs/1401.4452} {\path{arXiv:1401.4452}},
  \href {http://dx.doi.org/10.1142/S0217751X14300117}
  {\path{doi:10.1142/S0217751X14300117}}.

\bibitem{2013morrisdietz}
J.~A. {Dietz}, T.~R. {Morris}, {Asymptotic safety in the f(R) approximation},
  Journal of High Energy Physics 1 (2013) 108.
\newblock \href {http://arxiv.org/abs/1211.0955} {\path{arXiv:1211.0955}},
  \href {http://dx.doi.org/10.1007/JHEP01(2013)108}
  {\path{doi:10.1007/JHEP01(2013)108}}.

\bibitem{2015demsau}
M.~{Demmel}, F.~{Saueressig}, O.~{Zanusso}, {A proper fixed functional for
  four-dimensional Quantum Einstein Gravity}, Journal of High Energy Physics 8
  (2015) 113.
\newblock \href {http://arxiv.org/abs/1504.07656} {\path{arXiv:1504.07656}},
  \href {http://dx.doi.org/10.1007/JHEP08(2015)113}
  {\path{doi:10.1007/JHEP08(2015)113}}.

\bibitem{2006PhLB:capoz}
S.~{Capozziello}, S.~{Nojiri}, S.~D. {Odintsov}, A.~{Troisi}, {Cosmological
  viability of -gravity as an ideal fluid and its compatibility with a matter
  dominated phase}, Physics Letters B 639 (2006) 135--143.
\newblock \href {http://arxiv.org/abs/astro-ph/0604431}
  {\path{arXiv:astro-ph/0604431}}, \href
  {http://dx.doi.org/10.1016/j.physletb.2006.06.034}
  {\path{doi:10.1016/j.physletb.2006.06.034}}.

\bibitem{defe}
A.~{de Felice}, S.~{Tsujikawa}, {f(R) Theories}, Living Reviews in Relativity
  13 (2010) 3.
\newblock \href {http://arxiv.org/abs/1002.4928} {\path{arXiv:1002.4928}},
  \href {http://dx.doi.org/10.12942/lrr-2010-3}
  {\path{doi:10.12942/lrr-2010-3}}.

\bibitem{cosa15}
A.~{Codello}, J.~{Joergensen}, F.~{Sannino}, O.~{Svendsen}, {Marginally
  deformed Starobinsky gravity}, Journal of High Energy Physics 2 (2015) 50.
\newblock \href {http://arxiv.org/abs/1404.3558} {\path{arXiv:1404.3558}},
  \href {http://dx.doi.org/10.1007/JHEP02(2015)050}
  {\path{doi:10.1007/JHEP02(2015)050}}.

\bibitem{Finelli:2016cyd}
F.~Finelli, et~al., {Exploring Cosmic Origins with CORE: Inflation}\href
  {http://arxiv.org/abs/1612.08270} {\path{arXiv:1612.08270}}.

\bibitem{Matsumura:2013aja}
T.~Matsumura, et~al., {Mission design of LiteBIRD}[J. Low. Temp.
  Phys.176,733(2014)].
\newblock \href {http://arxiv.org/abs/1311.2847} {\path{arXiv:1311.2847}},
  \href {http://dx.doi.org/10.1007/s10909-013-0996-1}
  {\path{doi:10.1007/s10909-013-0996-1}}.

\bibitem{Kogut:2011xw}
A.~Kogut, et~al., {The Primordial Inflation Explorer (PIXIE): A Nulling
  Polarimeter for Cosmic Microwave Background Observations}, JCAP 1107 (2011)
  025.
\newblock \href {http://arxiv.org/abs/1105.2044} {\path{arXiv:1105.2044}},
  \href {http://dx.doi.org/10.1088/1475-7516/2011/07/025}
  {\path{doi:10.1088/1475-7516/2011/07/025}}.

\bibitem{Abazajian:2016yjj}
K.~N. Abazajian, et~al., {CMB-S4 Science Book, First Edition}\href
  {http://arxiv.org/abs/1610.02743} {\path{arXiv:1610.02743}}.

\bibitem{Bonanno:2017gji}
A.~Bonanno, S.~J. Gabriele~Gionti, A.~Platania, {Bouncing and emergent
  cosmologies from Arnowitt-Deser-Misner RG flows}, Class. Quant. Grav. 35~(6)
  (2018) 065004.
\newblock \href {http://arxiv.org/abs/1710.06317} {\path{arXiv:1710.06317}},
  \href {http://dx.doi.org/10.1088/1361-6382/aaa535}
  {\path{doi:10.1088/1361-6382/aaa535}}.

\end{thebibliography}

\end{document}